\documentclass{article}
\usepackage[table,xcdraw]{xcolor}
 \usepackage{graphicx} 
\usepackage[left=1in,right=1in,top=1in,bottom=1in]{geometry}
\usepackage{amsmath}
\usepackage{amssymb}
\usepackage{amsfonts}
\usepackage{bm}
\usepackage{multirow}
\usepackage{float}
\usepackage{tikz}
\usepackage{fp}
\usetikzlibrary{shapes, arrows, positioning, automata, arrows.meta, quotes}
\usepackage{mathrsfs}
\usepackage{breqn}
\usepackage{hyperref}
\usepackage{subfig}
\usepackage{comment}
\usepackage[section]{placeins}
\newcommand{\pos}{\ensuremath{\mathcal{I}}}
\newcommand{\posp}{\ensuremath{\mathcal{I}'}}
\newcommand{\vac}{\ensuremath{\mathcal{V}}}
\newcommand{\vacp}{\ensuremath{\mathcal{V}'}}
\usepackage{fancyhdr}

\usepackage{listofitems}
\newcommand{\TM}[8]{%
\FPeval\resulta{clip(1-#1-#2)}%
\FPeval\resultb{clip(1-#3-#4-#7)}%
\FPeval\resultc{clip(1-#5-#6-#8)}%
\ensuremath
{\begin{bmatrix}
\resulta & 0 & 0 & #7 & #8\\ 
#1 & 0 & 0 & #3 & #5\\
#2 & 0 & 0 & #4 & #6\\
0 & 1 & 0 & \resultb & 0\\
0 & 0 & 1 & 0 & \resultc
\end{bmatrix}}}
\title{Probabilistic Modeling of Antibody Kinetics Post Infection and Vaccination: A Markov Chain Approach
}
\author{ Rayanne A. Luke$^{1,2,*,\dag}$ $\cdot$ Prajakta Bedekar$^{2,\dag}$ $\cdot$ Lyndsey M. Muehling$^3$ \\
Glenda Canderan$^3$ $\cdot$ Yesun Lee$^4$ $\cdot$ Wesley A. Cheng$^4$  $\cdot$ Judith A. Woodfolk$^3$ \\
Jeffrey M. Wilson$^3$ $\cdot$ Pia S. Pannaraj$^4$ $\cdot$ Anthony J. Kearsley$^2$ \\
\scriptsize $^1$Department of Mathematical Sciences, George Mason University, Fairfax, Virginia, 22030, USA \\
\scriptsize $^2$Information Technology Laboratory, National Institute of Standards and Technology, Gaithersburg, Maryland, 20899, USA \\
\scriptsize $^3$Division of Asthma, Allergy and Immunology, University of Virginia, Charlottesville, Virginia, 22903, USA \\
\scriptsize $^4$Department of Pediatrics, University of California San Diego, La Jolla, California, 92093, USA \\
\scriptsize $^*$Corresponding author: rluke@gmu.edu.
\scriptsize $^\dag$These authors contributed equally to this work.
}
\date{\today}

\begin{document}

\maketitle

\begin{abstract}
    
Understanding dynamics of  antibody levels is crucial for characterizing time-dependent response to immune events: either infections or vaccinations. The sequence and timing of these events significantly influence antibody level changes. Despite extensive interest in the recent years and many experimental studies, the effect of immune event sequences on antibody levels is not well understood. Moreover, disease or vaccination prevalence in the population are time-dependent. This, alongside the complexities of personal antibody kinetics, makes it arduous to analyze a sample immune measurement from a population. A rigorous mathematical characterization can inform public health decision making.

 \textbf{Relevance to Life Sciences.}  A key result of this paper is an antibody response modeling framework for an arbitrary number of multiclass immune events--the first of its kind to the best of our knowledge. Our model is ideal for characterizing immune event sequences, referred to as personal trajectories. To illustrate our ideas, we apply our mathematical framework to longitudinal severe acute respiratory syndrome coronavirus 2 (SARS-CoV-2) data from individuals with multiple documented infection and vaccination events. This approach is fully generalizable to other diseases that exhibit waning immunity, such as influenza, respiratory syncytial virus (RSV), and pertussis. Our work is an important step towards a comprehensive understanding of antibody kinetics for infectious diseases that could lead to an effective way to analyze the protective power of natural immunity or vaccination, predict missed immune events at an individual level, and inform booster timing recommendations.
 
 \textbf{Mathematical Content.} 
 We design a rigorous mathematical characterization in terms of a time-inhomogeneous Markov chain model for event-to-event transitions coupled with a probabilistic framework for the post-event antibody kinetics of multiple immune events. Probabilistic models appropriately describe these measurements as they capture the natural
variability in a population’s antibody response. We build probability density models for population response since the emergence of a disease via a discrete convolution of immune state transmission probabilities and personal response models, repeatedly invoking the definition of conditional probability and the law of total probability. Importantly, our coupled framework simultaneously tracks immune state and antibody response. This novel modeling approach surpasses the susceptible-infected-recovered (SIR) characterizations by rigorously tracing the probability distribution of population antibody response across time.
\end{abstract}

\section{Introduction}

Understanding antibody kinetics is crucial for developing effective vaccination strategies and predicting the spread of diseases. Antibody kinetics describe the dynamic antibody response to immune events for an individual. When an emergent disease has been circulating for some time and vaccines are introduced, we must consider situations with multiple infections and vaccinations per individual. Even if single-event antibody responses are known, it is unclear how the personal response to subsequent events will combine. 
Potential avenues of investigation include assuming the immune response mounts from its current distribution due to a subsequent immune event, and assuming that the acceleration of immune response is potentially modulated by the antibody level already present in the blood \cite{srivastava2024sars}. 
Experimentally, significant prior work investigated longitudinal immune response dynamics (e.g., \cite{diep2023successive,guo2023durability,liu2023persistence}).

While understanding individual antibody kinetics is essential, it is equally important to analyze the population-level antibody distribution to inform public health decisions. Random population sampling for antibody testing and analyzing such test information can yield valuable guidance for making population-level decisions 
\cite{caini2020meta,peeling2020serology}.
The population-level  antibody distribution is governed by the prevalence for disease versus vaccination. Many seroprevalence studies have been conducted (e.g., \cite{osborne2000ten,pollan2020prevalence,bajema2021estimated}), and a few mathematical approaches to seroprevalence analysis were developed, including estimation of the force of infection \cite{hoze2025rsero}. 
The complex  interplay of different factors affecting population antibody distributions suggests great value in developing a coherent mathematical framework to analyze measurements in support of decision-making. 
 No prior  mathematical theory has simultaneously addressed (i) antibody levels, (ii) prevalence, (iii) multiple classes, (iv) time-dependence, and (v) multiple immune events, nor their complex multi-scale interactions, to the best of our knowledge. In one example topics (iii) and (iv) were analyzed, but the effects of each event was studied separately. In this work, the authors applied models based on exponential and power-law kinetics to the post-peak immune vaccination response, studying several types of antibodies and investigating distributions for halving times and model parameters  \cite{murphy2024understanding}. Standard models for disease transmission, such as susceptible-infected-recovered (SIR) models or statistical regression models, can be employed to study topics (ii)--(v) but cannot track population-level antibody responses across time 
(e.g., \cite{d2020assessment,mcmahon2020reinfection,quick2021regression,roberto2021sars}). In \cite{dick2021covid}, the authors designed an age-structured model for boosting and waning immunity from infection or vaccination to severe acute respiratory syndrome coronavirus 2 (SARS-CoV-2) infection, thus addressing topics (ii)-(v). In another transmission model example, a network model was designed to characterize the spread of multiple contagions through a population (ii), (iii) \cite{stanoev2014modeling}.  
A recent within-host differential equation model investigated reinfection (v) and considered the increased immune capacity  induced by an additional infection with a variant \cite{schuh2024mathematical}; such models ignore population-level trends. One research group \cite{hay2019characterising} used Markov chain Monte Carlo methods to study multiple influenza infections or vaccinations in ferrets, addressing   antibody levels (i) and time-dependence (iv).   Limited theory has attempted to characterize the antibody response over time to infection and/or vaccination beyond our prior work \cite{bedekar2022prevalence, bedekar2025prevalence}; in the latter we studied the setting in which individuals can be infected or vaccinated but not both, and designed an unbiased prevalence estimation scheme via transition probability matrices.

Our aim in this paper is to address prevalence and time-dependence in a multiclass and multi-event setting. We begin by constructing the general framework of the problem in    Section \ref{sec:response_models}. Next, we introduce personal response models for antibody response after an arbitrary number of immune events in   Section \ref{sec:pers_response_models}.  We then present a two-event time-inhomogeneous model (Section \ref{sec:two_event_model}), expanding the number of allowed classes from the five in our previous work \cite{bedekar2022prevalence} to thirteen. The corresponding analysis is assembled in parallel to our previous work.  
 The aforementioned model provides insight into the most general time-inhomogeneous model, which is introduced in  Section  \ref{sec:full_time_inhom}. 
The corresponding conditional probabilities can be written in terms of the time-dependent antibody responses and the probability of a particular immune event sequence. 
 This is the most mathematically complete model, but corresponding estimations can be intractable in the presence of low amounts of data. However, provided the incidence rates change slowly or have stabilized, we can utilize a time-homogeneous model to approximate the transmission.
  In  Section \ref{subsec:time_homog_model}, we consider a time-homogeneous model with an unlimited number of allowable immune events, with incidence rates assumed to be constant over time.
  Next, in    Section \ref{sec:results_data} we demonstrate the results of  Section  \ref{sec:two_event_model} numerically by building antibody response models for multiple events using data from two separate cohorts reporting multiple immune events, first reported in \cite{keshavarz2022trajectory}, \cite{canderan2025distinct}, and \cite{congrave2022twelve}\footnote{Certain commercial equipment, instruments, software, or material are identified in this paper in order to specify the experimental procedure adequately. Such identification is not intended to imply recommendation or endorsement by the National Institute of Standards and Technology, nor is it intended to imply that the materials or equipment identified are necessarily the best available for the purpose.}. 
    Our Section \ref{sec:MC_simulations} demonstrates the results of    Section \ref{subsec:time_homog_model} numerically by simulating population transmission.
 The discussion includes further analysis of the reduced time-inhomogeneous model, limitations, extensions, and implications for immunologists (Section \ref{sec:disc}). Section \ref{sec:relevant_math_background} includes relevant mathematical background for non-mathematicians.

\section{Modeling paradigm: broad ideas}
\label{sec:response_models}

We study antibody kinetics in a population as a disease spreads, with implications for public health policy decisions. 
Our goal is to assign meaning to a set of antibody measurements randomly sampled from a population on a given day. Below, we list relevant terms and definitions from applied diagnostics. In the next subsection, we describe notation specific to this paper.

\begin{itemize}
    \item The na{\"i}ve class consists of individuals who have no history of infection or vaccination. Such individuals are often referred to as ‘negative’ in a binary classification setting.

\item The infected class consists of individuals whose most recent immune event is infection.  Such individuals are often referred to as ‘positive’ in a binary classification setting.

\item The vaccinated class consists of individuals whose most recent immune event is immunization.

\item Incidence refers to the fraction of new infections or vaccinations in the total population during a given time step \cite{bouter2023textbook}. We define an infection incidence and a vaccination incidence.

\item A class prevalence during a given time step after the emergence of a disease is the fraction of individuals in the population in that class on that time step.

 \item Personal timeline refers to the duration since infection or vaccination for an individual.
 
 \item Absolute timeline denotes time relative to the emergence of the disease.
\end{itemize}

\subsection{Modeling framework}
\label{sec:modeling_framework}

For all subsequent sections, we assume the following holds true. A blood 
sample from an individual is measured to obtain an antibody measurement $\bm{r}$, a vector in some compact domain $\Omega \subset \mathbb{R}^n$. The boundaries of $\Omega$ are governed by the measurement range of the instrument used. We use $t$ to indicate time in the personal timeline, which is the duration since infection or vaccination for an individual. We use $T$ to denote time in the absolute timeline in the emergence of the disease.
Antibody measurements are usually reported in regular, binned time intervals; therefore, we consider time to be discrete throughout the manuscript.

We use the blueprint from our prior work \cite{bedekar2025prevalence}  that was constructed for single-event models, but we relax this assumption to allow for multiple immune events. We employ a Markov chain approach to model the antibody kinetics of an emergent disease.
Antibody levels depend on transitions into infection or vaccination states. Thus, the models for population-level antibody response should depend on a weighted sum of all potential transitions, which can be represented via a transition matrix. The population-level antibody response over time can thus be formulated in terms of the transition probabilities weighted by personal antibody response evolution. Given a current state and time $t$ in personal timeline, one can compute the transition probability for the next time step. This motivates a discrete-time Markov chain framework, because only the current state and conditions ($\bm{r},t,T$) affect the next state.

The states we consider are na{\"i}ve ($N$), newly infected ($\pos$), newly vaccinated ($\vac$), previously infected ($\posp$), and previously vaccinated ($\vacp$). In the next section, we expand from these five states to thirteen to count secondary immune events as explicit, separate events.
Each state has an associated prevalence: $q_N(T ),  q_{\pos} (T ), q_{\vac}(T), q_{\posp}(T)$, and $q_{\vacp} (T )$, denoting the fraction of the population in each state at time step $T$. Prevalence quantifies the total fraction of the population incident into that state so far and thus takes values in the range $[0, 1]$. 

A related quantity is the incidence rate, or fraction of the total population that newly moves to a particular state at time step $T$ of the absolute timeline; these quantities also take values in the range $[0, 1]$. The incidences describe the movement from a particular state to another; $f_{\pos N}$ describes the incidence from the na{\"i}ve state to the newly infected state.
Among the incidence rates, notice that $f_{\pos N}(T),$ for example, means that this is the fraction of the population as a whole who moved from na{\"i}ve in  the previous step $T-1$ to newly infected now at time $T$. In other words, $f_{\pos N}+f_{\pos \posp}+f_{\pos \vacp} = f_{\pos}$. This $f_{\pos}(T)$ represents the fraction of the population that has moved into the newly infected class on time step $T$.

A transition matrix $S$ contains the probabilities of moving between states. Here, $S(i, j)$ is the probability of moving to state $i$ \textit{from} state $j$. Let $T = 0$ index the emergence of a disease. We assume that our initial state vector is $X_1 = \bm{e}_1$ to model the disease emergence, so that all the members of the population are in state $N$ with probability $1$ on the day before the disease emerges. Here, $\bm{e}_1$ is the first unit vector. This assumption simplifies our model: we characterize an emergent disease rather than an endemic virus. Let $X_j$ denote the state, or class, at time step $j$.  Denote the transition probabilities by $s$. 

An ultimate goal of our work is to track antibody response and immune state simultaneously. Probabilistic models are an appropriate way to describe these measurements because they capture the natural variability in a population’s antibody response. Such a model can predict the probability of observing an antibody measurement $\bm{r}$ at time $T$ in the absolute timeline, given that the measurement corresponds to an individual in a particular state (e.g, previously infected, $\posp$). We refer to these absolute timeline models as conditional probabilities.
To build the conditional probabilities, we combine transition probabilities with \textit{personal response models}. The personal response models depend on $t$, the time in the personal timeline relative to infection or vaccination, and are introduced in the next section.

\section{Personal response models}
\label{sec:pers_response_models}

Personal antibody responses are governed by the immune events experienced by the individual. The propagation of a disease or vaccination thereof through a population over time affects the number of people with a particular antibody response. For an individual, we assume that the response is solely governed by their immune events. Moreover, the antibody distributions for previously infected and vaccinated individuals change with time: after infection or vaccination, an individual's antibody response starts na{\"i}ve, reaches a peak, and decays back to na{\"i}ve over long periods of time. Thus, we model antibody response at a given time for these immune events, with antibody level $r \in \mathbb{R}$ for the sake of simplicity. To create a probability distribution, we select a parameterized model that qualitatively describes the population. Examples of distributions commonly used to model biological phenomena include the beta, exponential, gamma, normal, and uniform. We choose to use gamma distributions to model these antibody responses due to their infinite divisibility property. A random variable that follows a gamma distribution with shape $\alpha$ and scale $\beta$ has the following probability density for antibody level $r$,
\begin{equation}
    \text{Gamma}(r; \alpha, \beta) = \frac{1}{\Gamma(\alpha) \beta^{\alpha}} r^{\alpha-1} e^{-r/\beta},
    \label{eq:pdf_gamma}
\end{equation}
where $\Gamma$ is the gamma function.
The na{\"i}ve antibody distribution does not change with time and we thus model it with constant shape and scale parameters with the corresponding density,
\begin{equation}
    N(r) = \text{Gamma}(r; \alpha_N, \beta_N).
    \label{eq:naive_model}
\end{equation}
We make a modeling choice to prescribe time-dependence to the shape parameter of the gamma distribution to introduce the effect of vaccination or infection. Specifically, we extend $\alpha$ to vary in time for both the vaccinated and infected responses as
\begin{equation}
    \alpha_{1,z}(t) = \frac{\theta_{z} t}{1 + \phi_{z} t^{k_z}} + \alpha_N,\ \text{ where } z \in \{\pos,\vac\}.
    \label{eq:shape1_new}
\end{equation}
 Plotted against $t$, the shape parameter $\alpha$ increases to a peak and then slowly decays, just like the antibody response to infection or vaccination.
Here, $z$ denotes either an infected or vaccinated response; note that the responses are to particular \emph{events} rather than to a \emph{state} as the time-dependent immune response mounts after an immune event, either infection or vaccination. The subscript 1 on $\alpha$ denotes that this is a single event model. The parameter $\theta_z>0$ governs the magnitude of the antibody response. The parameter $\phi_z>0$ determines the time after the immune event when a peak response is achieved. The parameter $k_z > 1$ determines the decay rate of this response, with a larger decay rate corresponding to a less durable immune response and a faster return to na{\"i}ve antibody distribution. For instance, if an immune event generates a quick, high magnitude, durable response, it corresponds to a model with a large $\theta$, a large $\phi$, and a small $k$. Notice that $\alpha_{1,z}(0) = \lim\limits_{t \rightarrow \infty} \alpha_{1,z}(t) = \alpha_N,$ ensuring that the immune response is identical to the na{\"i}ve response at the time of immune event and at large times if no other immune events have taken place. The Gamma scale parameter  $\beta$  is maintained as the na{\"i}ve distribution value  $\beta_N$. 

A comment on our choice of densities is in order. Note that the sum of independent  gamma distributions with identical scale parameters is a gamma distribution whose shape parameter is the sum of the individual shapes. That is, letting $X_i \sim g(k_i,\theta)$ where $k_i,\theta$ are shape and scale respectively, if all $X_i$ are independent, then $\sum_{i=1}^n X_i \sim g(\sum_{i=1}^n k_i,\theta),$ where $g$ denotes a gamma-distributed random variable. Notice that other families of distributions such as normal and chi-squared also possess similar characteristics, termed as `infinite divisibility,' and could make good potential candidates for modeling if such a choice is supported by aspects of available data. As we  assume that the \emph{additional} antibody response mounted by an individual post-event is independent of the response till that time, we are able to construct an additive shape model that can be generalized to multiple events.

Let $R_z(r, t)$ denote the probability density of observing a measurement of $r$ at $t$ time steps after infection or vaccination. The model for personal timeline response is then given by 
\begin{equation}
    R_z(r,t) := \text{Gamma}(r; \alpha_{1,z}(t), \beta_N).
    \label{eq:single_event_model}
\end{equation}
Each slice of the model $R_z(r,t)$ in personal timeline $t$ is a probability distribution; $R_z(r,t)$ evolves in time. 

To extend our single event model  \eqref{eq:shape1_new}, \eqref{eq:single_event_model} to a two-event model, we introduce a modified shape term
\begin{equation}
\alpha_{2,\{z_1,z_2\}}(t, \tau; \theta_{z_1}, \theta_{z_2}, \phi_{z_1}, \phi_{z_2}, k_{z_1}, k_{z_2}) := \frac{\theta_{z_1} \tau}{1+\phi_{z_1} \tau^{k_{z_1}}} + \frac{\theta_{z_2} (t - \tau)}{1+\phi_{z_2} (t- \tau)^{k_{z_2}}} + \alpha_N,
\label{eq:shape_2event_k}
\end{equation}
where $t$ remains the elapsed time between the first event and the day of measurement, $\tau$ is the gap between event 1 and event 2 (relative days between the two events), and $t-\tau$ is the time between event 2 and the day of measurement.  The parameters $\theta_{z_1}, \phi_{z_1},$ and $k_{z_1}$ are taken as the optimal parameters from maximum likelihood estimation (MLE) for the single event model, while $\theta_{z_2}, \phi_{z_2}$, and $k_{z_2}$ are obtained from MLE for individuals with a second event with their respective $\tau$. The model assumes that the immune response due to the second immune event at any time $t>\tau$ mounts \emph{in addition to} the immune response generated by the first immune event until time $\tau$. Thus, the personal timeline model for two events is given by
\begin{equation}
    R_{\{z_1,z_2\}}(r,t, \tau) := \text{Gamma}(r; \alpha_{2,\{z_1,z_2\}}(t, \tau; \theta_{z_1}, \theta_{z_2}, \phi_{z_1}, \phi_{z_2}, k_{z_1}, k_{z_2}), \beta_N). 
    \label{eq:2_model}
\end{equation}

 We generalize the assumption from \cite{bedekar2022prevalence} for a single event to two to require that $R_{\{\posp,\posp\}}(r, t, t) = R_{\{\posp,\vacp\}}(r, t, t) = R_\posp(r, t-1)$ and $R_{\{\vacp,\vacp\}}(r, t, t) = R_{\{\vacp,\posp\}}(r, t, t) = R_\vacp(r, t-1)$; that is, when $t - \tau = 0$, meaning the second event has just occurred, the two event models are identical to the single event versions.

We now extend our model to consider multiple events, significantly increasing its usefulness; this situation is frequently encountered in real-life scenarios. Let $\{z_m\}$ be the sequence of $\mathcal{M}$ events for an individual, with $\boldsymbol{\tau}$  the vector of corresponding time gaps between subsequent events. Then, the modified shape function is
\begin{equation}
\alpha_{\mathcal{M},\{z_m\}}(t, \boldsymbol{\tau}; \theta_{\{z_m\}},\phi_{\{z_m\}}, k_{\{z_m\}}) := \sum\limits_{j=1}^{\mathcal{M}-1}\frac{\theta_{z_j} \tau_j}{1+\phi_{z_j} \tau_j^{k_{z_j}}} + \frac{\theta_{z_{\mathcal{M}}} (t - \sum \tau_j)}{1+\phi_{z_{\mathcal{M}}} (t- \tau)^{k_{z_{\mathcal{M}}}}} + \alpha_N,
\label{eq:shape_Mevents}
\end{equation}
with the corresponding response,
\begin{equation}
    R_{\{z_m\}}(r,t, \boldsymbol{\tau}) := \text{Gamma}(r; \alpha_{\mathcal{M},\{z_m\}}(t, \boldsymbol{\tau}; \theta_{\{z_m\}},\phi_{\{z_m\}}, k_{\{z_m\}}), \beta_N). 
    \label{eq:M_model}
\end{equation}
Notice that for every additional immune event, we must only estimate three additional parameters, $\{\theta_{z_{\mathcal{M}}}\}$, $\{\phi_{z_{\mathcal{M}}}\}$, $\{k_{z_{\mathcal{M}}}\}$, improving the usability of the model. As before, the $\theta$ and $\phi$ values for the first $\mathcal{M}-1$ events are already determined by the MLE for the previous models. If data dictates a consistent decay rate following a particular immune event, we will require that $\{k_{z_{\mathcal{M}}}\}$  take on only one of two values, one for infection, $k_{\pos}$, and the other for vaccination, $k_{\vac}$. We emphasize that this general response model is the convolution of the responses for the sequence of events for an individual. Now, we assess two different models to study the transmission of infection and vaccination through the population and how their responses convolve with said transmission to generate the probability density of the population response on an absolute timeline. 

\section{Two-event time-inhomogeneous model}  
\label{sec:two_event_model}

\begin{figure}[h]
    \centering
    \includegraphics[scale=1.2]{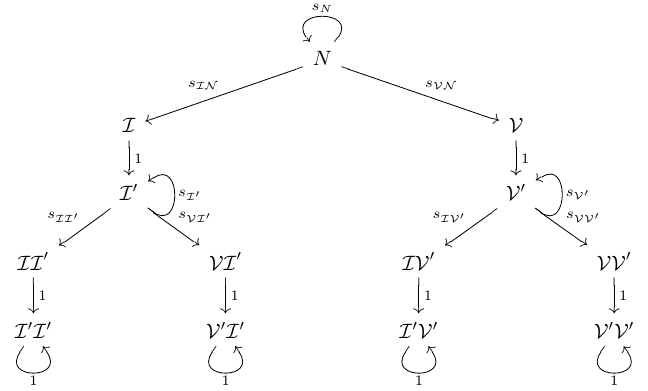}
    \caption{Graph of a model where two events are allowable: two infections, two vaccinations, infection after vaccination, or vice versa.}
    \label{fig:two_infect_vax}
\end{figure}

In this section, we present a model in which two immune events are allowable: two infections, two vaccinations, or one after the other. This is a generalization of our work in \cite{bedekar2025prevalence} in the sense that to incorporate multiple events, we consider new states instead of letting individuals pass between the original five states. The states that we consider for the two-event transmission model described in this section are na{\"i}ve ($N$), newly infected ($\pos$), newly vaccinated ($\vac$), previously infected once ($\posp$), previously vaccinated once ($\vacp$), newly infected with a second infection ($\pos \posp$), newly vaccinated after infection ($\vac \posp$), newly infected after vaccination ($\pos \vacp$), newly vaccinated with a second vaccination ($\vac \vacp$), previously infected twice ($\posp \posp$), previously infected and vaccinated ($\posp \vacp$), previously vaccinated and infected ($\vacp \posp$), and previously vaccinated twice ($\vacp \vacp$). We employ a graph to represent our framework, in which each state or class is a node and transitions between classes are directed, weighted edges. We let $s_N (T )$ denote the weight of the degenerate edge to $N$ from $N$, or the probability of staying na{\"i}ve. The probabilities $s_{\pos N} (T )$ and $s_{\vac N} (T )$ weight the edges to $\pos$ from $N$ and to $\vac$ from $N$, indicating infection or vaccination, respectively; the other transition probabilities have analogous interpretations. Figure  \ref{fig:two_infect_vax} shows the allowable movements between the thirteen states. Note, 
once an individual has experienced two immune events, they move into either a previously infected or previously vaccinated state with probability $1$ and then stay there indefinitely. In this model, we also forbid explicit movement back to the na{\"i}ve state after an immune event has occurred; doing so assumes that the time scale of the problem is short.

\subsection{Transition probabilities}

By definition, the transition probabilities depend on the incidence rates. At time step $T$, the fraction of the population in a particular state that becomes newly infected or vaccinated is divided by the relative size of that particular state population at the previous step to obtain the transition probability.  We thus have
\begin{subequations}
\begin{align}
s_{\pos N}(T) &= \frac{f_{\pos}(T)}{q_N(T-1)}, \qquad
s_{\vac N}(T) = \frac{f_{\vac}(T)}{q_N(T-1)}, \label{eq:from_N} \\
s_{\pos \posp}(T) &= \frac{f_{\pos \posp}(T)}{q_{\posp}(T-1)}, \qquad
s_{\vac \posp}(T) = \frac{f_{\vac \posp}(T)}{q_{\posp}(T-1)}, \qquad
s_{\pos \vacp}(T) = \frac{f_{\pos \vacp}(T)}{q_{\vacp}(T-1)}, \qquad
s_{\vac \vacp}(T) = \frac{f_{\vac \vacp}(T)}{q_{\vacp}(T-1)} \label{eq:second_event}.
\end{align}
\label{eq:trans_prob_two_event}
\end{subequations}
Here, we have grouped the transition probabilities as follows: \eqref{eq:from_N} transitions out of the na{\"i}ve state and \eqref{eq:second_event} transitions to a new immune event from a previous immune event state.
By a total probability argument, we find the remaining, self-edge weights as
\begin{subequations}
\begin{align}
    s_N(T) & =  1 - s_{\pos N}(T) - s_{\vac N}(T)\\
    s_\posp(T) & =  1 - s_{\pos \posp}(T) - s_{\vac\posp}(T)\\
    s_\vacp(T) & =  1 - s_{\vac \vacp}(T) - s_{\pos\vacp}(T) .
\end{align}
\end{subequations}

We define $q_N(-1) = 1$ to be consistent with our assumption that everyone is in the na{\"i}ve state on the day before the disease emerges. When $q_{(\star)}(T - 1) = 0$ for any state $(\star)$, there are no individuals currently in that state so transitions out of that state are impossible. In such a situation we define the corresponding transition probabilities to be zero; that is, those involving $q_{(\star)}(T - 1)$ in the denominator of their definition.

The transition matrix for movement from time step $T-1$ to time step $T$ is given by \eqref{eq:two_event_matrix}. Here, the ordering of the matrix rows and columns is $N, \pos, \vac, \posp, \vacp, \pos \posp, \vac \posp, \pos \vacp, \vac \vacp, \posp \posp, \vacp \posp, \posp \vacp, \vacp \vacp$. In   Figure \ref{fig:two_infect_vax}, this corresponds to an ordering of top to bottom and left to right within a level.
\setcounter{MaxMatrixCols}{13}
\begin{equation}
    S(T) = \begin{bmatrix}
s_N(T) & 0 & 0 & 0 & 0 & 0 & 0 & 0 & 0 & 0 & 0 & 0 & 0 \\ s_{\pos N}(T) & 0 & 0 & 0 & 0 & 0 & 0 & 0 & 0 & 0 & 0 & 0 & 0 \\
s_{\vac N}(T) & 0 & 0 & 0 & 0 & 0 & 0 & 0 & 0 & 0 & 0 & 0 & 0 \\
0 & 1 & 0  & s_{\posp}(T) & 0 & 0 & 0 & 0 & 0 & 0 & 0 & 0 & 0 \\
0 & 0 & 1 & 0 & s_{\vacp}(T)  & 0 & 0 & 0 & 0 & 0 & 0 & 0 & 0  \\
0 & 0 & 0 & s_{\pos \posp}(T) & 0 & 0 & 0 & 0 & 0 & 0 & 0 & 0 & 0 \\
0 & 0 & 0 & s_{\vac \posp}(T) & 0 & 0 & 0 & 0 & 0 & 0 & 0 & 0  & 0 \\
0 & 0 & 0 & 0 & s_{\pos \vacp}(T) & 0 & 0 & 0 & 0 & 0 & 0 & 0 & 0  \\
0 & 0 & 0 & 0 & s_{\vac \vacp}(T) & 0 & 0 & 0 & 0 & 0 & 0 & 0 & 0  \\
0 & 0 & 0 & 0 & 0 & 1 & 0 & 0 & 0 &  1 & 0 & 0 & 0  \\
0 & 0 & 0 & 0 & 0 & 0 & 1 & 0 &  0 & 0 & 1 & 0 & 0 \\
0 & 0 & 0 & 0 & 0 & 0 & 0 & 1 & 0 & 0 & 0 & 1 & 0  \\
0 & 0 & 0 & 0 & 0 & 0 & 0 & 0 & 1 & 0 & 0 &  0 & 1 
    \end{bmatrix}
    \label{eq:two_event_matrix}
\end{equation}

Multi-step transitions involving the matrix $S$ can be represented in terms of the entries of the matrix-vector product of the subsequent transitions. A $(\sigma +1)$-step transition from time step 0 to time step $\sigma$ is represented by $H_{\sigma}$, where
\begin{equation}
    H_{\sigma} = S(\sigma) S(\sigma - 1) \cdots S(1) S(0) = \left( \prod_{t = 0}^{\sigma} S(\sigma - t) \right).
    \label{eq:H_transition}
\end{equation}
Here, $\sigma - t$ is used to enforce indexing of the product in the correct order. Assuming everyone starts in the na{\"i}ve class on time step 0, $H_\sigma \mathbf{e}_1$ is a $(13\times 1)$ vector with the $j^{th}$ entry corresponding to the probability that an arbitrary individual in the population ends up in the $j^{th}$ state in $\sigma$ time steps. The matrix multiplication to obtain $H_{\sigma}$ accounts for every sequence that ends in a particular immune event. See  Appendix \ref{sec:appendix_model} for explicit calculations; specifically \eqref{eq:state_prob_full}.
As an example, consider the following,
\begin{equation}
 \text{Prob}(X_T = \posp) = \sum_{t = 0}^{T-1} \left( \prod_{\sigma = t+2}^T s_{\posp}(\sigma) \right) s_{\pos N}(t)  \prod_{\ell = 0}^{t-1} s_N(\ell).
 \label{eq:posp_state_prob}
 \end{equation}
 $\text{Prob}(X_T = \posp)$ is partitioned by the day of infection, which can be any day $t$ from 0 to $T-1$; this movement is indicated by the transition probability $s_{\pos N}(t)$. Here, one is na{\"i}ve before the day of infection, moves with probability 1 to the previously infected state $\posp$ on the next day ($t+1$), then stays previously infected thereafter up to and including day $T$. 
Note that $\text{Prob}(X_T = N) = q_N(T)$; that is, the na{\"i}ve state probability equals the corresponding prevalence. Similarly, $\text{Prob}(X_T = \pos) = f_{\pos}(T)$ and $\text{Prob}(X_T = \vac) = f_{\vac}(T)$, so the single new infection and single new vaccination state probabilities are equal to the corresponding incidences.

 \subsection{Conditional probabilities}
 \label{sec:cond_prob_two_event}

 The goal of considering a two-event model is that the trajectories to all states are explicit, following one branch of the graph only, and thus the conditional probability densities can be expressed in a straightforward manner because all possible paths can be easily enumerated. We present the conditional probabilities from the top ``level'' of  Figure \ref{fig:two_infect_vax} to the bottom.

The models for single new infections and vaccinations  remain the same as the formulations given in \cite{bedekar2025prevalence}. The expressions are simple because there is only one possible sequence of state transitions: $N N \cdots N \pos$, where the transition from $N$ to $\pos$ occurs on time step $T$. The conditional probability density for an antibody measurement $\bm{r}$ during timestep $T$ in the absolute timeline given that the sample comes from a newly (singly) infected individual is given by
\begin{equation}
    \text{Prob}(X_T = \pos) = \text{Prob}(\bm{r}, T | X_{T-1} = N, X_T = \pos) = R(\bm{r}, 0) = N(\bm{r}),
\end{equation}
and similarly,  $\text{Prob}(X_T = \vac) = N(\bm{r})$. These are the two states in the second ``level'' of the graph in  Figure \ref{fig:two_infect_vax}.

Now we compute Prob$(\bm{r}, T | X_T = \posp)$ and Prob$(\bm{r}, T | X_T = \vacp)$, the probabilities of previous (single) infection and previous (single) vaccination. These are the two states in the third ``level'' of the graph in   Figure \ref{fig:two_infect_vax}. Since the set of previously (once) infected individuals can be partitioned by the day on which they were infected (once), we can use the law of total probability and then the definition of conditional probability to rewrite the conditional probability as
\begin{equation}
\begin{split}
\text{Prob}(\bm{r}, T | X_T = \posp) & = \sum_{t = 0}^{T-1} \text{Prob}(\bm{r}, T, X_t = \pos | X_T = \posp) \\
& = \frac{1}{\text{Prob}(X_T = \posp)} \sum_{t = 0}^{T-1} \text{Prob}(\bm{r}, T, X_t  = \pos, X_T = \posp).
    \end{split}
\end{equation}
We then compute
\begin{equation}
\begin{split}
\text{Prob}(\bm{r}, T, X_t  & = \pos, X_T = \posp)  = R_{\pos}(\bm{r}, T-t) \underbrace{s_{\pos N}(t) \prod_{\ell = 0}^{t-1} s_N(\ell)}_{\text{Prob}(X_t = \pos) = \langle H_t \bm{e}_1, \bm{e}_2 \rangle} \prod_{\sigma = t + 2}^T s_{\posp}(\sigma) , \\
& = R_{\pos}(\bm{r}, T-t) \langle H_t, \bm{e}_1, \bm{e}_2 \rangle \prod_{ \sigma = t + 2}^T s_{\posp}(\sigma) ,
\end{split}
\end{equation}
since $R_{\pos}(\bm{r}, T-t)$ gives the distribution of antibody response on time step $T$ in the absolute timeline for a person who was infected on day $t$. Thus, the conditional probability density is given by
\begin{equation}
    \text{Prob}(\bm{r}, T | X_T = \posp) = \frac{1}{\langle H_T \bm{e}_1, \bm{e}_4 \rangle} \left( \sum_{t = 0}^{T-1} R_{\pos}(\bm{r}, T-t) \langle H_t, \bm{e}_1, \bm{e}_2 \rangle  \prod_{ \sigma = t + 2}^T s_{\posp}(\sigma) \right).
\end{equation}
The expression for Prob$(\bm{r}, T | X_T = \vacp)$ is analogous. We note that this differs from the corresponding conditional probability in \cite{bedekar2025prevalence} by the inclusion of the product involving $s_{\posp}$; this is because in our prior work, $s_{\posp}(\sigma) = 1$ for all $\sigma$, as we forbid reinfections.

Next, we compute probabilities on the fourth ``level'' that indicate new, secondary immune events. We derive Prob$(\bm{r}, T | X_T = \pos \posp)$. Using the same approach as above, we have
\begin{equation}
\begin{split}
\text{Prob}(\bm{r}, T | X_T = \pos \posp) & = \sum_{t = 0}^{T-1} \text{Prob}(\bm{r}, T,  X_t = \pos | X_T = \pos \posp), \\
& = \frac{1}{\text{Prob}(X_T = \pos \posp)} \sum_{t = 0}^{T-1} \text{Prob}(\bm{r}, T, X_t = \pos, X_T = \pos \posp).
\end{split}
\end{equation}
We compute
\begin{equation}
\begin{split}
\text{Prob}(\bm{r}, T, X_t = \pos, X_T = \pos \posp) & = R_{\pos, \pos}(\bm{r}, T-t, T-t) s_{\pos \posp}(T) \underbrace{s_{\pos N}(t) \prod_{\ell = 0}^{t-1} s_N(\ell)}_{\text{Prob}(X_t = \pos) = \langle H_t \bm{e}_1, \bm{e}_2 \rangle} \prod_{\sigma = t + 2}^{T-1} s_{\posp}(\sigma) , \\
& = R_{\pos}(\bm{r}, T-t) \langle H_t \bm{e}_1, \bm{e}_2 \rangle s_{\pos \posp}(T) \prod_{\sigma = t + 2}^{T-1} s_{\posp}(\sigma),
\end{split}
\end{equation}
where we recall that the distribution of antibody response on the time step of secondary infection is the same as that as for the first infection up to that date. Together, this gives the conditional probability density as
\begin{equation}
   \text{Prob}(\bm{r}, T | X_T = \pos \posp) = \frac{s_{\pos \posp}(T)}{\langle H_T \bm{e}_1, \bm{e}_{10} \rangle} \sum_{t = 0}^{T-1} R(\bm{r}, T-t) \langle H_t, \bm{e}_1, \bm{e}_2 \rangle  \prod_{\sigma = t + 2}^{T-1} s_{\posp}(\sigma). 
\end{equation}

 Finally, we want to compute probabilities like Prob$(\bm{r}, T | X_T = \posp \posp)$, or the fifth ``level'' of   Figure \ref{fig:two_infect_vax}. Using the definition of conditional probability, this can be found by computing the state probabilities (done above) and probabilities of the form:
 \begin{equation}
     \text{Prob}(X_t = \pos, X_{\sigma} = \pos \posp, X_T = \posp \posp) = \underbrace{\left(s_{\pos N}(t) \prod_{\ell = 0}^{t-1} s_N(\ell) \right)}_{\text{Prob}(X_t = \pos) = \langle H_t \bm{e}_1, \bm{e}_2 \rangle}  s_{\pos \posp} (\sigma) \prod_{\ell = t + 2}^{\sigma - 1} s_{\posp} (\ell) .
 \end{equation}
 Then, since we have explicitly enumerated the possible trajectories to each state, we can express the desired conditional probability as a double sum as follows:
 \begin{equation}
 \label{eq:cond_prob_2event}
 \begin{split}
& \text{Prob}(\bm{r}, T | X_T = \posp \posp)  \\ 
& \qquad = \sum_{t = 0}^{T-3} \sum_{\sigma = t + 2}^{T-1} \text{Prob} (\bm{r}, T, X_t = \pos, X_{\sigma} = \pos \posp | X_T = \posp \posp), \\
& \qquad = \frac{1}{\text{Prob}(X_T = \posp \posp)} \sum_{t = 0}^{T-3} \sum_{\sigma = t + 2}^{T-1}  \text{Prob}(\bm{r}, T, X_t = \pos, X_{\sigma} = \pos \posp, X_T = \posp \posp), \\
& \qquad = \frac{1}{\text{Prob}(X_T = \posp \posp)} \sum_{t = 0}^{T-3} \sum_{\tau = 2}^{T-t-1}  R_{\pos, \pos}(\bm{r}, T-t, \tau)  \left(s_{\pos N}(t) \prod_{\ell = 0}^{t-1} s_N(\ell) \right)  s_{\pos \posp} (t + \tau) \prod_{\ell = t + 1}^{t + \tau - 1} s_{\posp} (\ell) , \\
& \qquad = \frac{1}{\langle H_T \bm{e}_1, \bm{e}_{10} \rangle} \sum_{t = 0}^{T-3} \sum_{\tau = 2}^{T-t-1} R_{\pos, \pos}(\bm{r}, T-t, \tau)  \langle H_t \bm{e}_1, \bm{e}_2 \rangle s_{\pos \posp} (t + \tau) \prod_{\ell = t + 1}^{t + \tau - 1} s_{\posp} (\ell),
\end{split}
\end{equation}
where we have changed notation slightly to use $\tau = \sigma - t$, the time gap between events 1 and 2, for consistency with the personal timeline models given by     \eqref{eq:shape_2event_k}-\eqref{eq:2_model}.

\section{General time-inhomogeneous model with five states}
\label{sec:full_time_inhom}

We now introduce the most general time-inhomogeneous model and provide insight into transmission probabilities and the conditional probability models that define the absolute timeline. This is a generalization of our work in \cite{bedekar2025prevalence} in the sense that to incorporate multiple events, we  consider new transitions (for instance, from previously infected to newly infected, newly vaccinated, na{\"i}ve) while keeping the Markov chain with the original five states.

We briefly consider the pros and cons of this general five-state framework versus the two-event, 13-state framework discussed in the previous section.  An advantage of the
general five-state approach is that it succinctly captures antibody response and immune state simultaneously; a disadvantage is that the corresponding conditional probability models are difficult to derive. On the other hand, the two-event, 13-state model is advantageous for the straightforward nature of its corresponding conditional probability models that are indexed by the two allowed immune events, but it is limited to short time scales or modeling non-circulating diseases because an extension to more events is cumbersome and computationally impractical.

\subsection{Transition probabilities}

\begin{figure}[h]
\centering
\includegraphics[width=0.5\linewidth]
{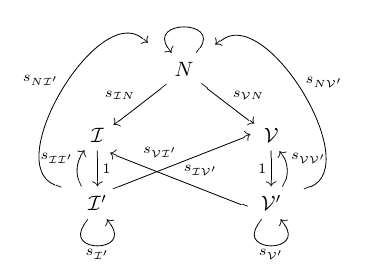}
\caption{Graph describing the allowable movements between states. Here, $N$ is na{\"i}ve, $\pos$ is newly infected, $\posp$ is previously infected,  $\vac$ is newly vaccinated, and $\vacp$ is previously vaccinated. Double subscripts on $s$ denote the transition probability from the second state to the first.}
        \label{fig:graph}
\end{figure}

The states that we consider are na{\"i}ve ($N$), newly infected ($\pos$), newly vaccinated ($\vac$), previously infected ($\posp$), and previously vaccinated ($\vacp$). 
The graph representing the transitions between states is shown in  Figure  \ref{fig:graph}. We let $s_N (T )$ denote the weight of the degenerate edge to $N$ from $N$, or the probability of staying na{\"i}ve. The probabilities $s_{\pos N} (T )$ and $s_{\vac N} (T )$ weight the edges to $\pos$ from $N$ and to $\vac$ from $N$, indicating infection or vaccination, respectively; the other transition probabilities have analogous interpretations. The only disallowed transitions are those that go from newly infected to any state other than previously infected, and the same for new vaccination.

 Many of the transition probabilities are calculated in the same way given by \eqref{eq:trans_prob_two_event}. We note the variations and new transition probabilities for this model:
\begin{subequations}
\begin{align}
s_{N \posp}(T) &= \frac{f_{N \posp}(T)}{q_{\posp}(T-1)}, \qquad
s_{N \vacp}(T) = \frac{f_{N \vacp}(T)}{q_{\vacp}(T-1)} \label{eq:return_to_N}, \\
s_\posp(T) & =  1 - s_{\pos \posp}(T) - s_{\vac\posp}(T) - s_{N \posp}(T)\\
    s_\vacp(T) & =  1 - s_{\vac \vacp}(T) - s_{\pos\vacp}(T) - s_{N \vacp}(T).
\end{align}
\label{eq:general_trans_prob}
\end{subequations}
Here, we have grouped the new transition probabilities as follows:  \eqref{eq:return_to_N} transitions back to the na{\"i}ve-like state from a previous immune event state, and the remaining are degenerate edge weights that now include the possibility of returning to the na{\"i}ve state.
  The transition matrix for movement from time step $T-1$ to time step $T$ is thus given by

    \begin{equation}
    \label{eq:trans_mat_1step}
    \begin{split}
    & S(T) \\
    & = \begin{bmatrix}
    1 - s_{\pos N}(T) - s_{\vac N}(T) & 0& 0 & \frac{f_{N \posp}(T)}{q_{\posp}(T-1)} & \frac{f_{N \vacp}(T)}{q_{\vacp}(T-1)} \\
     \frac{f_{\pos N}(T)}{q_N(T-1)}  & 0 & 0 & \frac{f_{\pos \posp}(T)}{q_{\posp}(T-1)} & \frac{f_{\pos \vacp}(T)}{q_{\vacp}(T-1)} \\
    \frac{f_{\vac N}(T)}{q_N(T-1)}  & 0 & 0 & \frac{f_{\vac \posp}(T)}{q_{\posp}(T-1)} &  \frac{f_{\vac \vacp}(T)}{q_{\vacp}(T-1)} \\
 0 & 1 & 0 & 1 - s_{\pos \posp}(T) - s_{\vac\posp}(T) -s_{N \posp}(T) & 0 \\
    0  & 0 & 1 & 0 & 1 - s_{\vac \vacp}(T) - s_{\pos\vacp}(T) - s_{N \vacp}(T)
        \end{bmatrix},
        \end{split}
    \end{equation}
        where the ordering is $N, \pos, \vac, \pos', \vac'$. 

We define the $(T+1)$-step transition matrix as 
\begin{equation} 
H_T = S(T) S(T-1)\cdots S(1)S(0) = \prod_{t=0}^T S(T-t).
\label{eq:H_transition_general}
\end{equation}
The quantities such as $\text{Prob}(X_T=N),\text{Prob}(X_T=\pos),$ and so on are defined as $\left<H_T \mathbf{e}_1,\mathbf{e}_j\right>$ for the appropriate index $j$ as in the previous work. We note that the population which is previously infected at time $T$, say, has myriad complex ways of reaching there, such as, being infected once, or reinfected multiple times, or vaccinated then infected, and so on. Especially in diseases where immunity conferred is not protective in the long run, the space of all possible event sequences can become quite large and cumbersome. The Markov chain notation helps keep this condensed by enumerating the sum of all such sequence probabilities in the entries of the $(T+1)$-step transition matrix.

Provided the transition probabilities in Figure \ref{fig:graph} are nonzero, we note that every state in this Markov chain can communicate with every other, that is, it is possible to go from any state to any other state in finitely many time steps. Moreover, the period of every state is $1$, making this Markov chain irreducible and aperiodic. By the Perron–Frobenius theorem, we know that a unique stationary distribution exists, i.e. there is a unique normalized eigenvector corresponding to the eigenvalue $1$, with nonnegative entries. For the transition matrix $S(T)$ from     \eqref{eq:trans_mat_1step}, $\mathbf{e_v}(T)$ is the stationary distribution.

\begin{equation}
\label{eq:stationary_dist_T}
\mathbf{v}(T) := \begin{bmatrix}
    s_{N\posp}(T)s_{N\vacp}(T) + s_{N\posp}(T) s_{\pos\vacp}(T) + s_{N\vacp}(T) s_{\vac\posp}(T)\\
    \left(1-s_\posp(T)\right)\left(s_{N\vacp}(T)s_{\pos N}(T) + s_{\pos \vacp}(T) s_{\pos N}(T) + s_{\pos\vacp}(T) s_{\vac N}(T)\right)\\\left(1-s_\vacp(T)\right)\left(s_{N\posp}(T)s_{\vac N}(T) + s_{\vac\posp}(T) s_{\pos N}(T) + s_{\vac \posp}(T) s_{\vac N}(T)\right)\\ s_{N\vacp}(T)s_{\pos N}(T) + s_{\pos \vacp}(T) s_{\pos N}(T) + s_{\pos\vacp}(T) s_{\vac N}(T)\\
    s_{N\posp}(T)s_{\vac N}(T) + s_{\vac\posp}(T) s_{\pos N}(T) + s_{\vac \posp}(T) s_{\vac N}(T)
\end{bmatrix}, \quad \mathbf{e_v}(T) := \frac{\mathbf{v}(T)}{\sum v_i(T)}.
\end{equation}
As a reminder, our order of states is $N, \pos,\vac,\posp,\vacp$. The stationary distribution denotes the fractional population distribution through the states so that action of the Markov chain doesn't change that distribution. With that in mind, notice that $v_2(T)=(1-s_\posp(T))v_4(T).$ As there are only two ways for an individual to move into the previously infected $\posp$ state: if they were newly infected (entered $\pos$, with relative fraction $v_2(T)$) or if they were in the previously infected state and didn't move out (stayed in \posp, relative fraction of $s_\posp(T) v_4(T)$. In other words, the fraction of population in the newly infected state $\pos$ for the stationary distribution has to make up for the fraction of the population that moves into a new state after $\posp$. Similar explanations exist for the other stationary distribution entries.

We ideally want to construct a \emph{limiting} distribution, i.e. a population distribution into different states in the limit of time. However, as this chain is time-inhomogeneous, whether there is a corresponding limiting distribution  will additionally depend on the time-dependent incidence rates; this is an intriguing mathematical question which we are exploring in ongoing research. We will analyze the limiting distribution for a time-homogeneous Markov chain in Section \ref{subsec:time_homog_model} and Section \ref{sec:MC_simulations} 
under the assumption that the incidence rates stabilize over time.

\subsection{Conditional probabilities}
\label{subsec:cond_prob_time_inhomogeneous}

The probability of observing antibody measurement $r$ at time $T$ given that the person is in the previously infected class at time $T$ is derived below. Notice that unlike in the previous work, a person can be infected, reinfected, vaccinated, revaccinated, and a mix of these before ending up at a final state. As a result, how a person reaches a state is just as important as the final state. Broadly, our conditional probability model is of the form,

\begin{align}
    \text{Prob}(\mathbf{r},T|X_T = \posp) & = \frac{1}{\text{Prob}(X_T = \posp)} \sum_{\left\{\{z_m\}_{m=0}^{\mathcal{M}}: z_{\mathcal{M}} = \pos\right\}} \text{Prob}(r,T,\{z_m\},X_T = \posp )\\
    & = \frac{1}{\left<H_T \mathbf{e}_1,\mathbf{e}_4\right>} \sum_{\left\{\{z_m\}_{m=0}^{\mathcal{M}}: z_{\mathcal{M}} = \pos\right\}} R_{\{z_m\}}(\bm{r},T, \boldsymbol{\tau}) \text{Prob}(\{z_m\}_{m=0}^{\mathcal{M}}, X_T = \posp).
\end{align}
Here, the summand is explicitly broken up into the product of the antibody response to a particular sequence and the probability of that sequence. Let $\mathbf{\Sigma}$ be the vector of absolute timeline times when the events happen, i.e., $\Sigma_i = \sum\limits_{j=1}^i \tau_j$. We can explicitly calculate the probability of a given sequence of events with a given $\boldsymbol{\tau}$,
\begin{equation}
    \text{Prob}(\{z_m\}_{m=0}^{\mathcal{M}}) = \left(\prod\limits_{i=0}^{\tau_1-1}s_{N}(i)\right) \left(s_{z_1N}(\tau_1) \prod\limits_{i=1}^{\mathcal{M}-1} s_{z_{i+1}\tilde{z}_i}(\Sigma_{i+1})\right) \left( \prod\limits_{i=1}^{\mathcal{M}}\prod\limits_{j=\Sigma_i+2}^{\Sigma_{i+1}-1} s_{\tilde{z}_i\tilde{z}_i}(j)\right),
    \label{eq:prob_seq_timeinhomogeneous}
\end{equation}
where, $\tilde{z}_i$ is the chronic version of the event $z_i$, i.e. if $z_i=\pos$, then $\tilde{z}_i=\posp$. In this calculation, the first set of parentheses contain the probability of staying in the na{\"i}ve class, the second set aggregates all transitions at events, and the third set aggregates all the probabilities where an individual stays in a given chronic state, $\posp$ or $\vacp$. We have made a simplifying assumption that $s_{N\posp} = s_{N\vacp} = 0$, which holds true for small $\mathcal{M}.$ In the long time cases, return to na{\"i}ve may have to be considered separately as an immune event. Notice that unlike our previous work, the conditional probabilities are now explicitly written in terms of $S(t)$ instead of $H(T)$ as every element in $H(T)$ comprises different sequence of events and corresponding antibody responses.

\subsection{Time-homogeneous model}
\label{subsec:time_homog_model}

The general, time-inhomogeneous model presented in Section \ref{sec:full_time_inhom} builds a Markov chain with desirable properties such as irreducibility and aperiodicity. However, due to the time-inhomogeneity, results on ergodicity are sparse. In this section, we thus present a corresponding time-homogeneous model with constant transition probabilities. This could help analyze situations where over time, the disease and vaccination incidence rates have stabilized. We again consider the states: na{\"i}ve ($N$), newly infected ($\pos$), newly vaccinated ($\vac$), previously infected ($\posp$), and previously vaccinated ($\vacp$). For the last two states, we assume that they are defined by the most recent immune event; if someone is infected and then later vaccinated, after the vaccination they belong to the previously vaccinated state.  

  The transition matrix for movement at any time is given by

    \begin{equation}
    \label{eq:trans_mat_1step_constant}
    S = \begin{bmatrix}
    1 - s_{\pos N} - s_{\vac N} & 0& 0 & s_{N \posp} & s_{N \vacp}\\
     s_{\pos N} & 0 & 0 & s_{\pos \posp}& s_{\pos \vacp} \\
    s_{\vac N}  & 0 & 0 & s_{\vac \posp} &  s_{\vac \vacp}\\
 0 & 1 & 0 & 1 - s_{\pos \posp} - s_{\vac\posp} -s_{N \posp} & 0 \\
    0  & 0 & 1 & 0 & 1 - s_{\vac \vacp} - s_{\pos\vacp} - s_{N \vacp}
        \end{bmatrix},
    \end{equation}

Modifying the probability calculated in     \eqref{eq:prob_seq_timeinhomogeneous} by removing the dependence of time and adding the possibility of repeated return to na{\"i}ve,  we obtain
\begin{equation}
    \text{Prob}(\{z_m\}_{m=0}^{\mathcal{M}}) = \left(s_{N}^{\tau_1}\right) \left(s_{z_1N} \prod\limits_{i=1}^{\mathcal{M}-1} s_{z_{i+1}\tilde{z}_i}\right) \left( \prod\limits_{i=1}^{\mathcal{M}} s_{\tilde{z}_i\tilde{z}_i}^{\tau_{i+1}-3}\right),
    \label{eq:prob_seq_time_homogeneous}
\end{equation}
where $z$ could be $N$ in addition to $\pos, \vac$ as earlier. If $z = N$, then $\tilde{z}=N$ as well.

Notice that this probability is dependent on the likelihood of particular transitions in that sequence. In particular, if the disease incidence has reduced and thereby the vaccination rates have reduced, then transition probabilities such as $s_{\pos\posp}, s_{\vac\posp},$ etc are low whereas $s_{\posp\posp},s_{\vacp\vacp},s_{N}$ are high. As a result, the probability of observing sequences with lower number of events goes up. That is, if the disease has begun to die down, the total immune events for an individual in the population are a lot lower on average. On the other hand, when a disease is on the rise, it is more likely for individuals to be infected/vaccinated, i.e., undergo immune events and therefore event sequences will be longer on average.

As previously stated, using the Perron-Frobenius theorem, there exists a unique stationary distribution for this time-homogeneous Markov chain, given by $\mathbf{e_v}$. 
\begin{equation}
\label{eq:stationary_dist}
\mathbf{v} := \begin{bmatrix}
    s_{N\posp}s_{N\vacp} + s_{N\posp} s_{\pos\vacp} + s_{N\vacp} s_{\vac\posp}\\
    \left(1-s_\posp\right)\left(s_{N\vacp}s_{\pos N} + s_{\pos \vacp} s_{\pos N} + s_{\pos\vacp} s_{\vac N}\right)\\\left(1-s_\vacp\right)\left(s_{N\posp}s_{\vac N} + s_{\vac\posp} s_{\pos N} + s_{\vac \posp} s_{\vac N}\right)\\ s_{N\vacp}s_{\pos N} + s_{\pos \vacp} s_{\pos N} + s_{\pos\vacp} s_{\vac N}\\
    s_{N\posp}s_{\vac N} + s_{\vac\posp} s_{\pos N} + s_{\vac \posp} s_{\vac N}
\end{bmatrix}, \quad \mathbf{e_v} := \frac{\mathbf{v}}{\sum v_i}.
\end{equation}

Additionally, this is also the \emph{limiting} distribution of the population. We explore this here and in Section \ref{sec:MC_simulations}. Ideally, the hope is that the fraction of the acutely infected population (in state $\pos$) is almost zero. From the equation above, notice that this is possible provided $s_\posp \approx 1$ and $s_{\pos N}, s_{\pos\vacp}$ remain low, i.e. most  previously infected stay in that state and are well protected, and incidence of infection is low.

\section{Example applied to SARS-CoV-2 antibody data}
\label{sec:results_data}

To validate our models, we use combined clinical data from \cite{canderan2025distinct} and \cite{keshavarz2022trajectory} and \cite{congrave2022twelve}, which we refer to as Datasets 1 and 2, respectively.
The data are spike immunoglobulin G (IgG) measurements taken: before any immune event, denoted by N as na{\"i}ve; after a single infection, denoted by I; after a two-dose vaccination sequence, denoted by V; after vaccination and a booster, denoted by VV (only for Dataset 1); and after a single vaccination dose following an infection, denoted by VI. Previously infected samples were taken from individuals whose infections were confirmed via reverse transcription polymerase chain reaction (RT-PCR) or antibody test, vaccinations were documented, and the na{\"i}ve samples were collected pre-pandemic, pre-vaccination, or confirmed negative via PCR or antibody test.   For Dataset 1, these values are considered together to have units of IU/mL, and for Dataset 2, the measurements are recorded in area under the curve (AUC) dimensionless units. The measurements from both datasets are log-transformed similarly to \cite{patrone2021classification, bedekar2022prevalence, luke2023optimal} to yield the unit-less, one-dimensional measurement
\begin{equation}
    r = \log_2(\tilde{r}+2)-1,
\end{equation} 
for Dataset 1 and 
\begin{equation}
    r = \log_2(\tilde{r}),
\end{equation} 
for Dataset 2, where the slight variation ensures that all transformed measurements remain nonnegative. All samples except na{\"i}ve have an antibody and personal time measurement pair $(r, t)$; na{\"i}ve samples have an antibody measurement $r$. Here, $t = 0$ indicates the day of the first immune event: either infection or the second dose of a two-dose vaccination sequence. To aid in the fitting procedure, we scale time by a characteristic unit of 100 days, so that we input $t = \tilde{t}/100$ into the optimizations.

\subsection{Dataset 1}
\label{sec:dataset_1}

\begin{figure}[h]
    \centering
\subfloat[][na{\"i}ve, infected, and infected then vaccinated]{\includegraphics[width=0.99\linewidth]{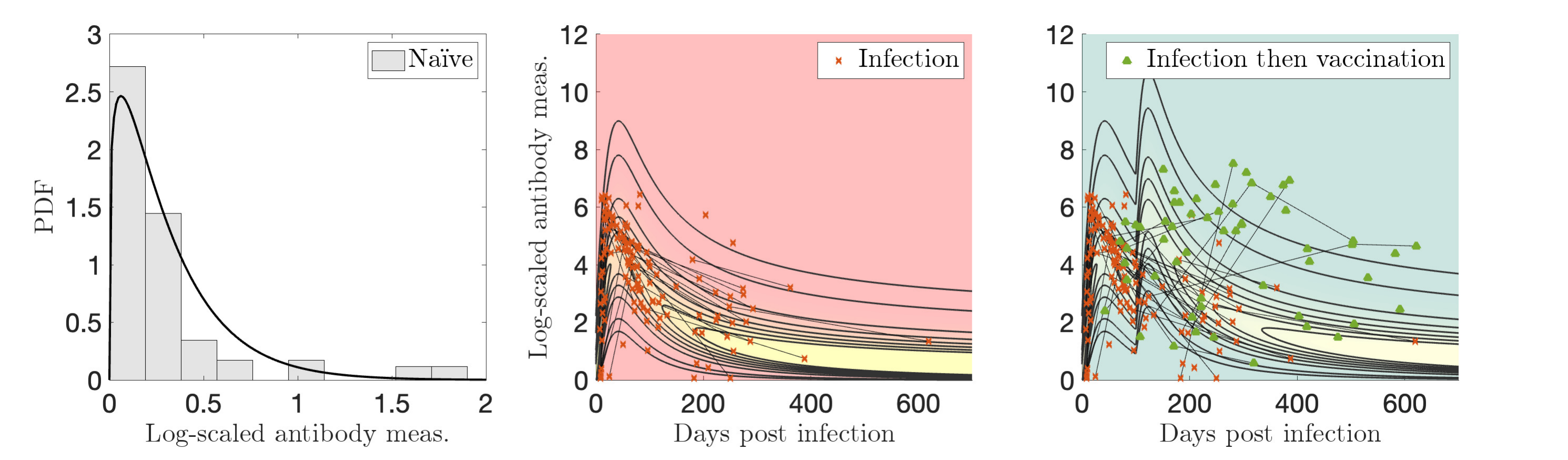} \label{fig:UVA_models_k_traj}} \\
\subfloat[][vaccinated and boosted]{\includegraphics[width=0.7\linewidth]{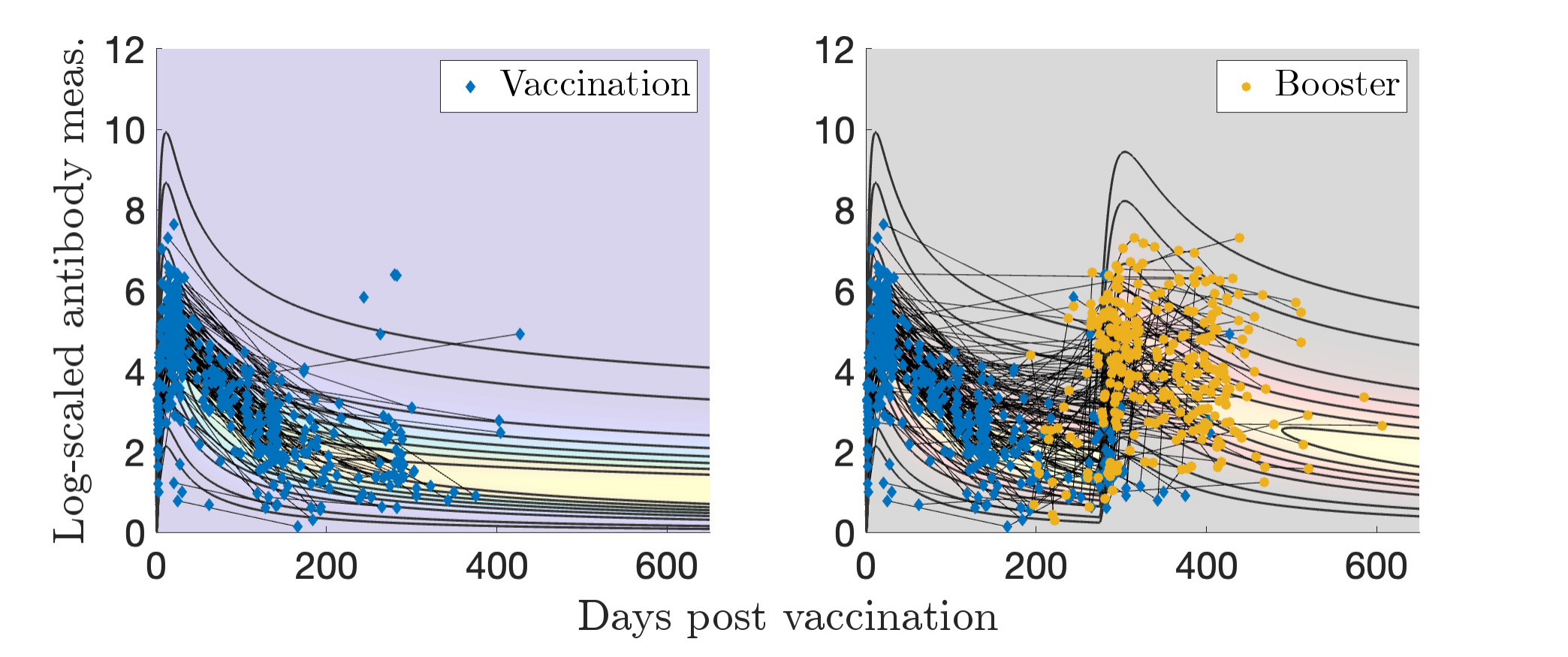} \label{fig:UVA_models_k_vax_traj}}
\caption{Log-transformed antibody measurements from Dataset 1 with corresponding probability models and personal trajectories (thin, solid black lines). (a)  na{\"i}ve, infected (I), and infected then vaccinated (VI) populations; (b) vaccinated (V) and boosted (VV) populations. In (a), PDF denotes the probability distribution function.}
    \label{fig:UVA_models_single}
\end{figure}

Measurement sample sizes for Dataset 1 are:  N ($n = 91$), I ($n = 130$), V ($n  = 453$), VI ($n = 56$), and VV ($n = 275$).  MLE applied to a gamma distribution for the na{\"i}ve sample measurements yields optimal parameters of  $\alpha_N = 1.23$ and $\beta_N = 0.256$. 
 We let the exponents vary in the optimization as $k_{\pos_1}, k_{\vac_1}$, or $k_{\vac_2}$ in \eqref{eq:shape_2event_k}. 
 The optimal parameter values are given in Table  \ref{table:MLE_params_UVA}. 
 Figure \ref{fig:UVA_models_single} shows the models for Dataset 1. The lightest region (yellow-ish in all plots) indicates higher probability, and the thick lines indicate probability contour levels from 0.001 (most outward) to 0.6 (most inward). The data are indicated with markers and the thin black lines indicate personal trajectories.
 Plotting the two-event model in two dimensions becomes difficult because there are two time variables involved, so we use  characteristic times between events of $\tau = 100$ days between infection and vaccination and $\tau = 275$ days for the booster visualization. These are all relatively short number of days between events compared to those recorded; the corresponding Supplementary Figure (video) shows all times between vaccination events. 

\begin{table}[h]
 \centering
\begin{tabular}{|l|r|r|r|r|r|r|r|r|r|r|}
\hline
\multicolumn{1}{|c|}{\textbf{Data}} & \multicolumn{1}{c|}{$\bm{\theta_{\pos_1} \textbf{or } \theta_{\vac_1}}$} & \multicolumn{1}{c|}{$\bm{\phi_{\pos_1} \textbf{or } \phi_{\vac_1}}$} & \multicolumn{1}{c|}{$\bm{k_{\pos_1} \textbf{or } k_{\vac_1}}$} & \multicolumn{1}{c|}{$\bm{\theta_{\vac_2}}$} & \multicolumn{1}{c|}{$\bm{\phi_{\vac_2}}$} & \multicolumn{1}{c|}{$\bm{k_{\vac_2}}$} \\ \hline
Infection & 79.0 & 5.63 & 2.10 &  &  &  \\ \hline
Vaccination & 520 & 51.4 & 1.51 &  &  &   \\ \hline
\begin{tabular}[c]{@{}l@{}} Infection, \\vaccination \end{tabular} & {\color[HTML]{9B9B9B} 79.0} & {\color[HTML]{9B9B9B} 5.63} & {\color[HTML]{9B9B9B} 2.10} & 101 & 14.3 & 1.93  \\ \hline
\begin{tabular}[c]{@{}l@{}}Vaccination, \\ booster \end{tabular} & {\color[HTML]{9B9B9B} 520} & {\color[HTML]{9B9B9B} 51.4} & {\color[HTML]{9B9B9B} 1.51} & 106 & 11.1 & 1.70  \\ \hline
\end{tabular}
\caption{Optimal parameters from maximum likelihood estimation for Dataset 1. Grayed out numbers indicate variables that are part of the model but are not part of the optimization for that data.
}
\label{table:MLE_params_UVA}
\end{table}


We quantify how well the models capture the data; this will validate our modeling approach as an appropriate one. We measure this in two ways: (1) how many subjects have all measurements from their personal trajectory contained within the model contours, i.e., probability $> 0.001$, and (2) how many subjects have more than half of their measurements with evaluated probabilities above 0.2, what we determine to be a reasonable ``high probability'' threshold. The results are summarized in Table \ref{table:contour_high_prob}.
For the single infection  model (I) and data in Figure  \ref{fig:UVA_models_k_traj}, the maximum probability observed is 0.84, and for the single vaccination model (V) and data in   Figure \ref{fig:UVA_models_k_vax_traj}, the maximum observed is 0.72. The results in Table \ref{table:contour_high_prob} validate our modeling procedure, showing that a significant portion of the data lie within the contours and for the vaccination data, have high probability evaluations. This may be expected, as vaccine dosage amounts are standardized in contrast to the variable viral load of infections, which may result in a tighter distribution of vaccine responses.

\begin{table}[h]
\centering
\begin{tabular}{|l|c|r|r|r|r|}
\hline
 & \textbf{Dataset} & \multicolumn{1}{c|}{\textbf{Infection}} & \multicolumn{1}{c|}{\textbf{Vaccination}} & \multicolumn{1}{c|}{\textbf{\begin{tabular}[c]{@{}c@{}}Infection then\\ vaccination\end{tabular}}} & \multicolumn{1}{c|}{\textbf{\begin{tabular}[c]{@{}c@{}}Vaccination, \\ booster\end{tabular}}} \\ \hline
\textbf{\begin{tabular}[c]{@{}l@{}}Pers. traj. in \\ contour lines\end{tabular}} & \multirow{2}{*}{1} & 66/82 (80 \%) & 177/251 (71 \%) & 9/14 (64 \%) & 113/118 (96 \%) \\ \cline{1-1} \cline{3-6} 
\textbf{\begin{tabular}[c]{@{}l@{}}Pers. traj. eval. \\ meas. $\bm{> 0.2}$\end{tabular}} &  & 49/82 (60 \%) & 122/251 (49 \%) & 3/14 (21 \%) & 81/118 (69 \%) \\ \hline
\textbf{\begin{tabular}[c]{@{}l@{}}Pers. traj. in\\ contour lines\end{tabular}} & \multirow{2}{*}{2} & 302/335 (90 \%) & 191/196 (97 \%) & 55/67 (82 \%) &  \\ \cline{1-1} \cline{3-6} 
\textbf{\begin{tabular}[c]{@{}l@{}}Pers. traj. eval. \\ meas. $\bm{> 0.2}$\end{tabular}} &  & 70/335 (21 \%) & 131/196 (67 \%) & 15/67 (22 \%) &  \\ \hline
\end{tabular}
\caption{Quantifying how well the models capture the data. Reported are how many subjects have all measurements from their personal trajectory contained within the model contours (probability $> 0.001$) and how many subjects have more than half of their measurements with evaluated probabilities above 0.2.}
\label{table:contour_high_prob}
\end{table}

 \subsection{Dataset 2}
 \label{sec:dataset_2}

\begin{figure}[h]
    \centering
    \includegraphics[width=1\linewidth]{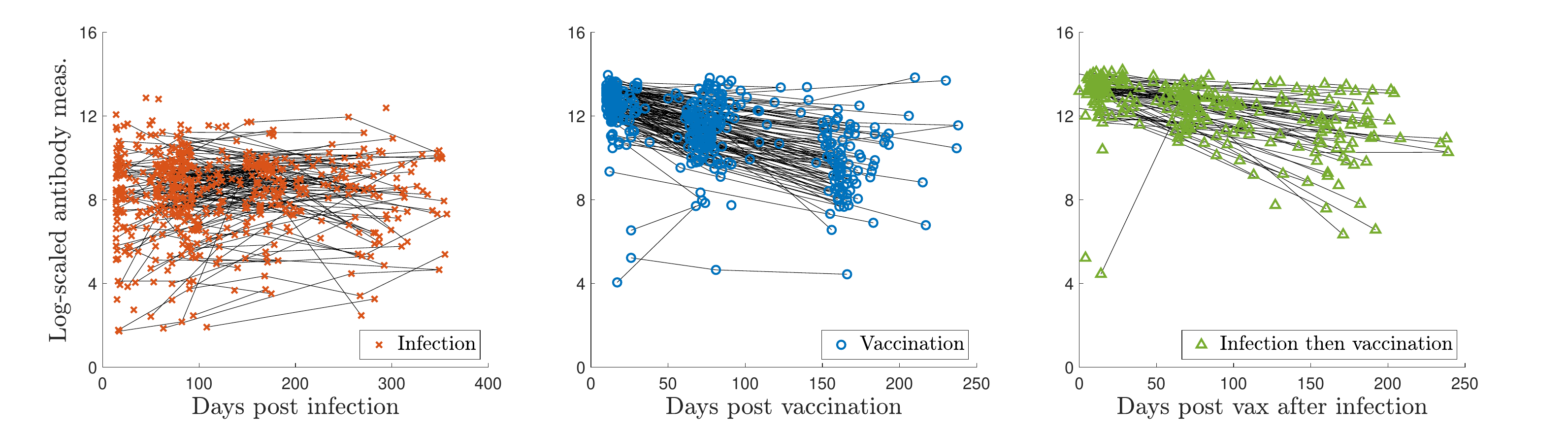}
      \caption{Non-titration-extrapolated, log-transformed antibody measurements  (markers) from the infected (I), vaccinated (V), and infected then vaccinated (VI) populations from Dataset 2 since the corresponding event plotted with personal trajectories (solid lines).}
    \label{fig:pers_traj}
\end{figure}

Measurement sample sizes for Dataset 2 are: I ($n = 562$), V ($n  = 410$), and VI ($n = 261$). The log-scaled but uncorrected data from Dataset 2 are shown in  Figure \ref{fig:pers_traj}. Notably, the characteristic decrease in antibody levels is missing for the infected samples. 
The AUC values from Dataset 2 are calculated using the optical density (OD) measurements from five serial dilutions (1:100, 1:300, 1:900, 1:2700, 1:8100). However, this maximal titration value may need to be raised because the OD values at a 1:8100 dilution were still nonzero and non-negligible for many samples. Thus, the actual infection and vaccination antibody levels, especially those recorded at early times, may be higher than  reported. Vaccination response tends to lead to higher measurements than infection, and the maximal titration value was selected based on the previously infected samples--further evidence that vaccination levels may actually be higher than those recorded. Thus, before validating our models on Dataset 2, we performed a titration-extrapolation of eligible measurements.

\begin{figure}[h]
    \centering
\subfloat[][Linear fit to meas. plotted against each other]{\includegraphics[width=0.38\linewidth]{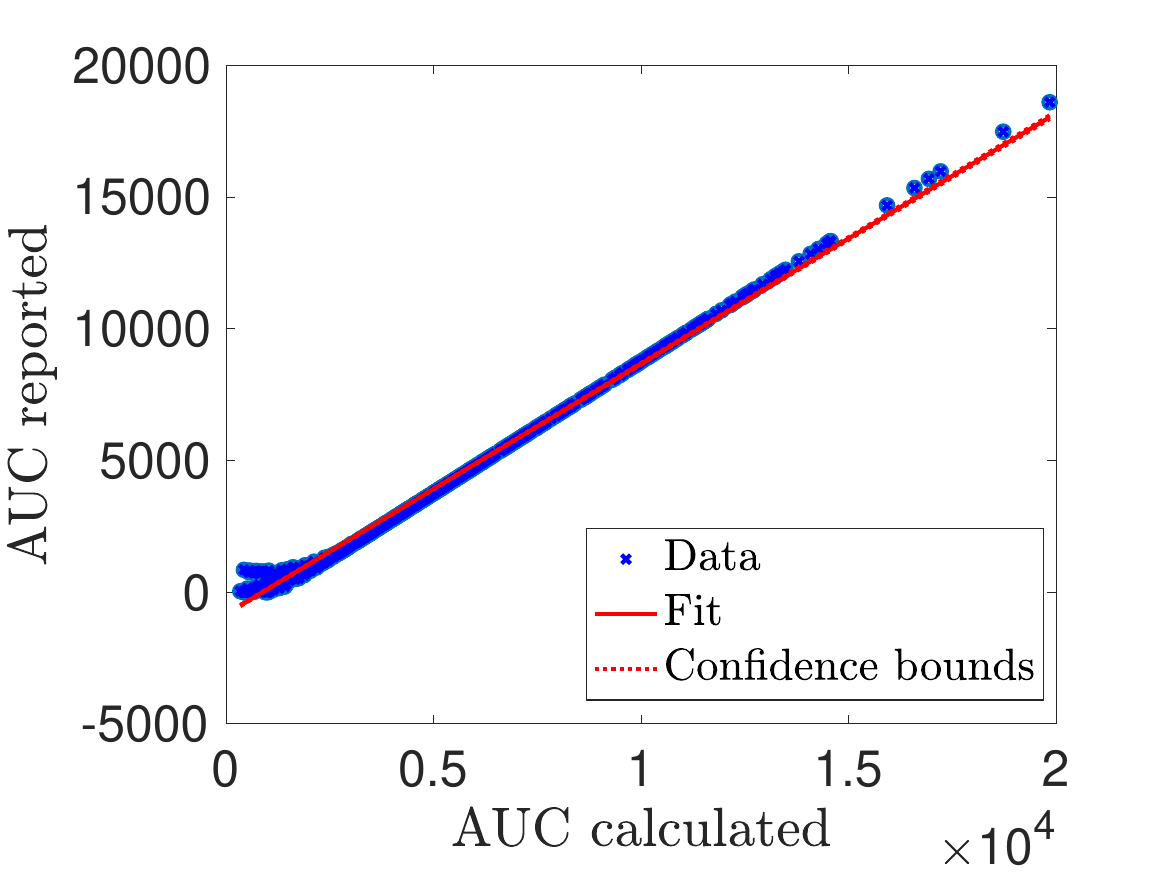}}
\subfloat[][All samples--dilution curves]{\includegraphics[width=0.38\linewidth]{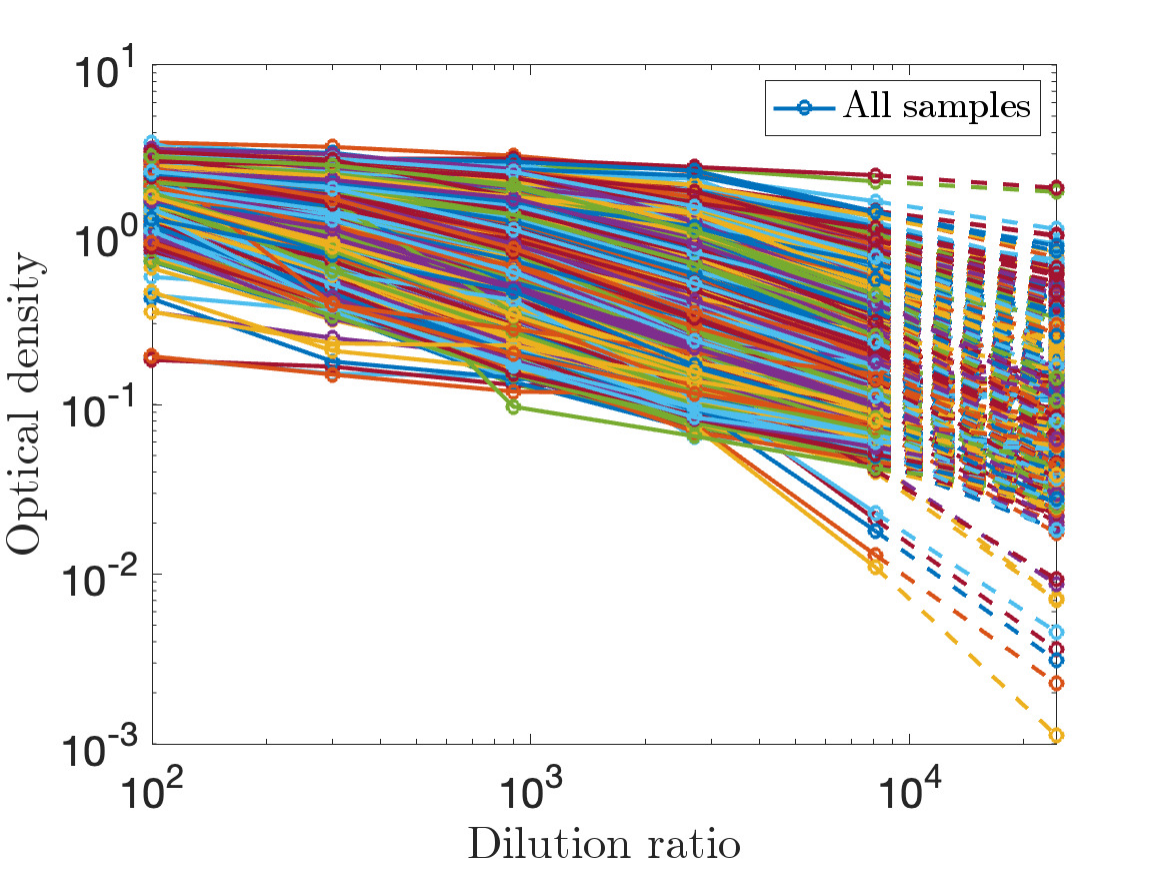}}
    \caption{For Dataset 2: (a) Linear fit to find intercept to translate our calculated AUC to that reported by \cite{congrave2022twelve}.  (b)  Titration curves with quadratic extrapolation with next titration level.}
    \label{fig:dil_curves2}
\end{figure}

\begin{figure}[h]
    \centering
    \includegraphics[width=\linewidth]{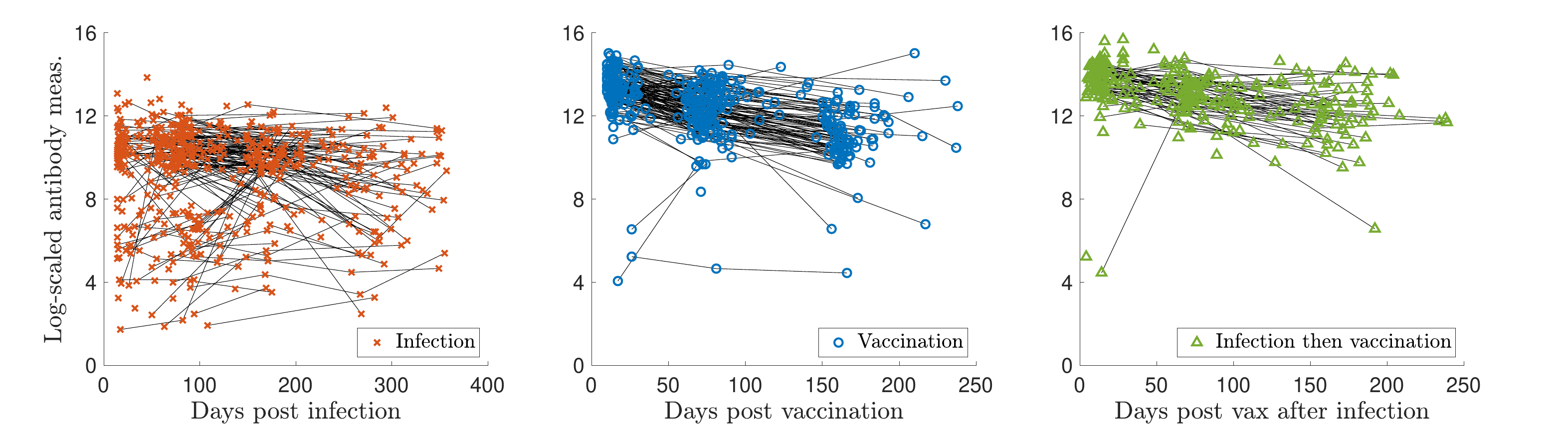}
      \caption{Log-transformed antibody measurements from the infected (I), vaccinated (V), and infected then vaccinated (VI) populations from Dataset 2 with personal trajectories. All AUC values with recorded and monotonically decreasing OD values have been titration-extrapolated using the next dilution level.}
    \label{fig:pers_traj_t2}
\end{figure}

 To adjust reported AUC measurements in Dataset 2, we received the optical density (OD) values corresponding to five serial dilutions (1:100, 1:300, 1:900, 1:2700, 1:8100) used to calculate the reported AUC \cite{congrave2022twelve}. We excluded samples without OD measurements, dilution curves of OD plotted against increasing dilution factor that were not monotonically decreasing, and negative OD values. After applying our exclusion criterion, 853/1244 samples were titration-extrapolated (68.6\%). The AUC is computed for the given dilutions via a trapezoidal rule. We plot the computed AUC  against the reported AUC in  Figure \ref{fig:dil_curves2}a. A linear relationship is observed for most samples, which is given by
\begin{equation}
    \rm{AUC}_{\rm reported} = 0.950 \  \rm{AUC}_{\rm calculated} - 818.
\end{equation}
This discrepancy may be due to a background level subtraction step during pre-processing.
Since the values seem to fall along a line of slope one with a negative intercept, the intercept of the best-fit line is recorded and used to shift the computed values to ``match'' the reported AUC.
Then, each dilution curve is extrapolated to the next titration level (1:24300) via a quadratic fit (dashed lines in  Figure \ref{fig:dil_curves2}b), and the AUC is re-calculated using a trapezoidal rule, as long as the extrapolated OD value is nonnegative. Figure  \ref{fig:pers_traj_t2} shows personal trajectories with extrapolated AUC values for all samples that met the inclusion criteria, broken down by class. On average, infection antibody levels decay more slowly than those of vaccinated individuals. We observe more of a decrease in antibody response for some infected trajectories post titration-extrapolation; many personal trajectories were fairly flat in the original data. We also observe a steeper decrease in antibody response for the vaccinated and vaccinated after infection measurements.


Dataset 2 did not include spike IgG measurements for na{\"i}ve samples, which provide an important baseline for our models, as \eqref{eq:single_event_model} builds off the parameters in \eqref{eq:naive_model}.
Therefore, for Dataset 2, we generate synthetic na{\"i}ve data with which to build the models for previously infected and vaccinated antibody responses $t$ days after infection or vaccination, and use $\alpha_n = 18.2$ and $\beta_n = 0.152$.  These parameters are chosen in such a way that they provide a reasonable guess for the distribution of previously infected and vaccinated responses at time zero post infection or inoculation. 
 The optimal parameters are shown in Table \ref{table:MLE_params_UCSD}.

\begin{figure}[h]
    \centering
\includegraphics[width=1\linewidth]{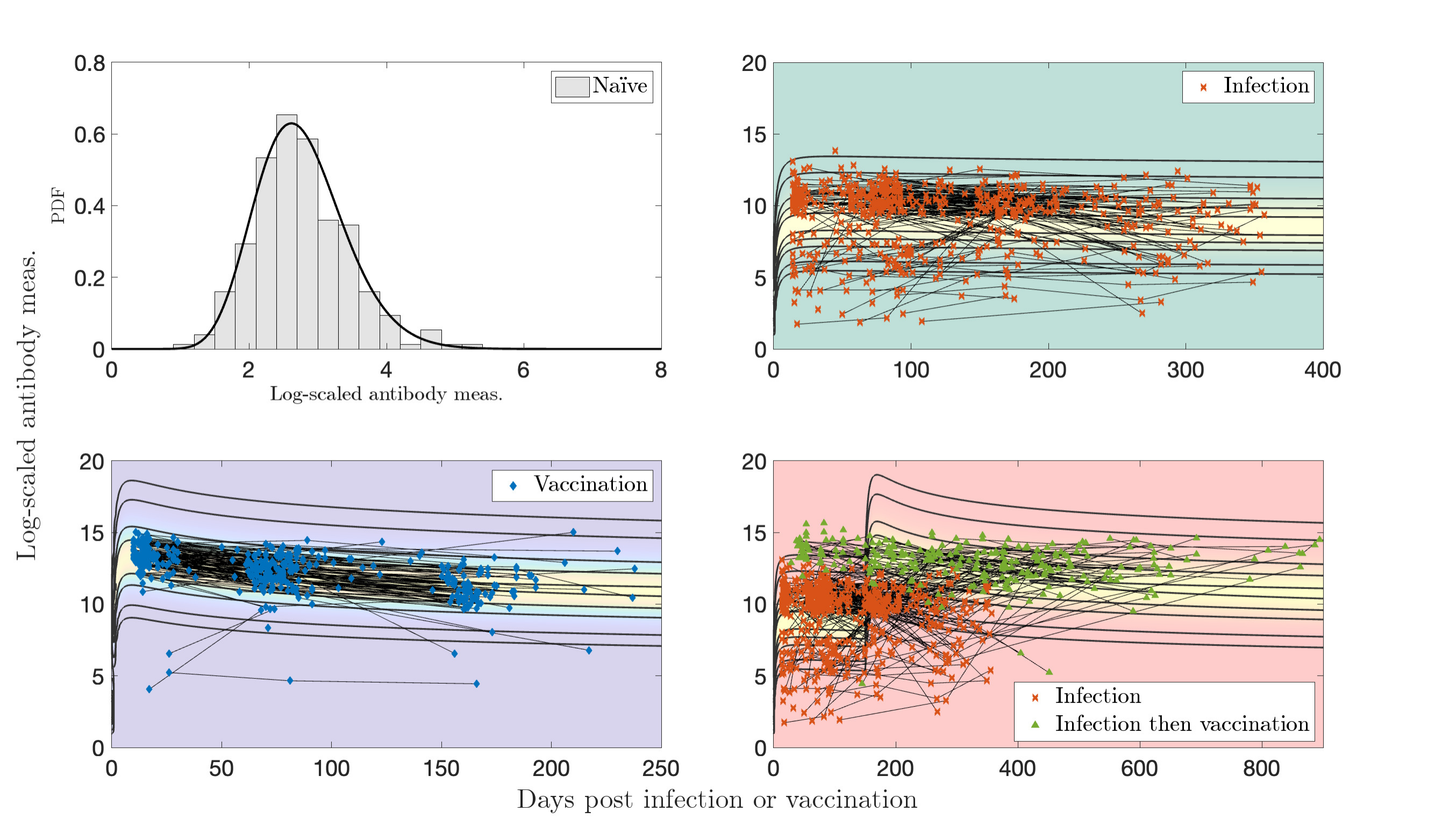}
    \caption{Log-transformed antibody measurements from the na{\"i}ve (synthetic), infected (I),  vaccinated (V), and infected then vaccinated (VI) populations from Dataset 2 with personal trajectories with corresponding probability models. Measurements with recorded monotonically decreasing OD values have been AUC-extrapolated using the next titration level. PDF denotes the probability distribution function.}
    \label{fig:CW_models_k_traj}
\end{figure}

Figure \ref{fig:CW_models_k_traj} shows the  models for Dataset 2. 
Similarly to Dataset 1, the lightest, yellow-ish region indicates higher probability, and the thick lines indicate probability contour levels from 0.001 (most outward) to 0.3 (most inward). The data are indicated with markers and the thin black lines indicate personal trajectories.
This plot is visualized using a relatively short period between events, $\tau = 150$ days.

The vaccine response peaks between 10 and 30 days post first dose, although the data is tightly clustered in time, as is expected from measurements obtained via a regimented clinical trial. The infection response does not exhibit a strong peak, even less so than the vaccine response and much less in comparison to Dataset 1. This may be due to different measurement systems and data pre-processing.

The vaccine response is more strongly clustered in the middle of the probability model, whereas a significant number of infected measurements show a lower than average response. This is to be expected, as asymptomatic infections, varying viral load, and difficulty in defining the index date of infection create additional variability in modeling immune response to infection.

 \begin{table}[H]
 \centering
\begin{tabular}{|l|r|r|r|r|r|r|r|}
\hline
\multicolumn{1}{|c|}{\textbf{Data}} & \multicolumn{1}{c|}{$\bm{\theta_{\pos_1} \textbf{or } \theta_{\vac_1}}$} & \multicolumn{1}{c|}{$\bm{\phi_{\pos_1} \textbf{or } \phi_{\vac_1}}$} & \multicolumn{1}{c|}{$\bm{k_{\pos_1} \textbf{or } k_{\vac_1}}$} & \multicolumn{1}{c|}{$\bm{\theta_{\vac_2}}$} & \multicolumn{1}{c|}{$\bm{\phi_{\vac_2}}$} & \multicolumn{1}{c|}{$\bm{k_{\vac_2}}$} \\ \hline
Infection & 2757 & 66.8 & 1.04 &  &  &   \\ \hline
Vaccination & 8113 & 135 & 1.11 &  &  &  \\ \hline
\begin{tabular}[c]{@{}l@{}} Infection, \\vaccination \end{tabular} & {\color[HTML]{9B9B9B} 2757} & {\color[HTML]{9B9B9B} 66.8} & {\color[HTML]{9B9B9B} 1.04} & 790 & 29.8 & 1.26  \\ \hline
\end{tabular}
\caption{Optimal parameters from maximum likelihood estimation for Dataset 2. Grayed out numbers indicate variables that are part of the model but are not part of the optimization for that data.
}
\label{table:MLE_params_UCSD}
\end{table}


Similarly to our analysis for Dataset 1, we quantify the goodness of fit of our modeling procedure in Table \ref{table:contour_high_prob}. For the single vaccination model (V), the maximum probability observed is 0.31, and for the single infection model (I), the maximum observed is 0.35. For the vaccination after infection model (VI), the maximum observed value is 0.36. The high percentages shown in the table validate our modeling procedure. Similar to Dataset 1, the vaccination data has especially high probability evaluations.

\section{Example applied to population transmission}
\label{sec:MC_simulations}

While the previous section was focused on antibody responses in individuals, the transmission of a disease through a population is another important factor that governs the total antibody distribution in a population over time. We now turn our attention to the different possible personal event sequences for individuals as an infection and its vaccination transmits through a population. We consider two separate time-homogeneous transition matrices and analyze their behavior over $50$ time steps, or around $2$ years. 

\begin{figure}[h]
    \centering
    \subfloat[][Histograms of immune status]{\includegraphics[width=0.9\textwidth]{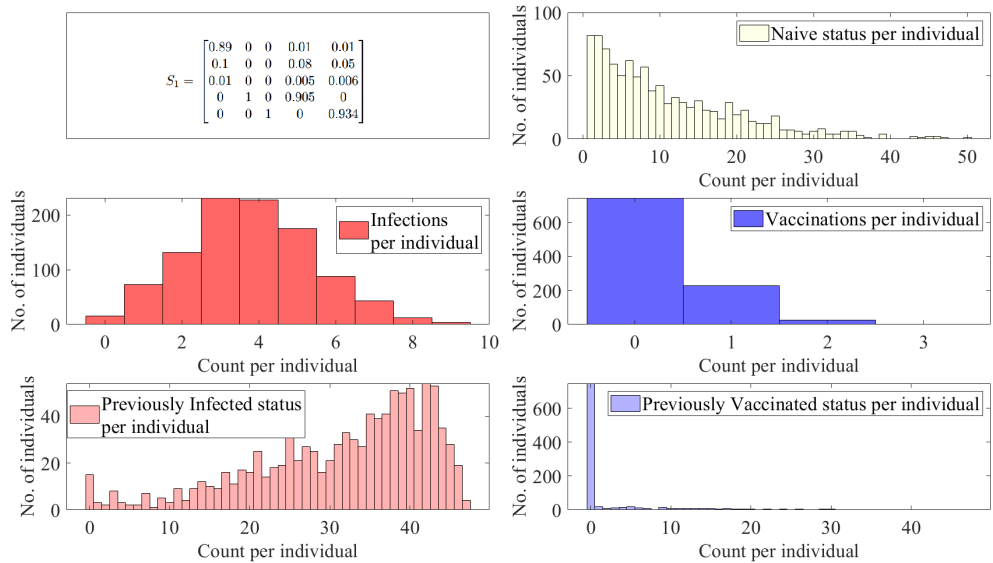} \label{fig:S1_counts}} \\
    \subfloat[][Three representative individual sequences]{\includegraphics[width=0.9\textwidth]{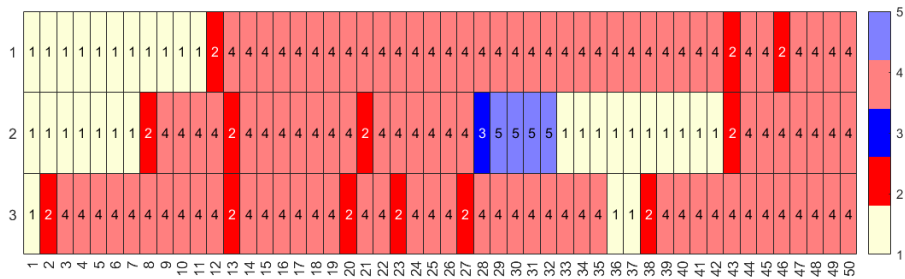} \label{fig:S1_sequences}}
    \caption{Simulations for a population with $1000$ individuals with a time-homogeneous transition matrix $S_1$, per $50$ time steps, i.e., approximately a two year period. (a) Histograms of counts of total na{\"i}ve, infections, vaccinations, previously infected status, and previously vaccinated status.
(b) Three  representative sequences. 
Light yellow denotes the na{\"i}ve state (1, $N$), red and pink  the newly and previously infected states (2, $\pos$ and 4, $\posp$), and blue and light blue the newly and previously vaccinated states (3, $\vac$ and 5, $\vacp$).}
\end{figure}

We first consider $S_1$ (Figure \ref{fig:S1_counts}) representing a disease spread with a rate of infection that is a magnitude higher than that of vaccination. Recall that our states are ordered as $N,\pos,\vac,\posp,\vacp$ in all transition probability matrices with the column representing the transition out of the corresponding state. We observe (Figure \ref{fig:S1_counts}) that due to larger probabilities of infection, the median number of infections per individual is $4$  and the median previously infected status is maintained for $35$ time steps. In contrast, the vaccinations have medians of $0$ and na{\"i}ve status is maintained for a median of $8$ time steps. Some representative individual event sequences are demonstrated in  Figure \ref{fig:S1_sequences}, each row represents a personal trajectory for an individual, the colors represent the immune status for the individual at the time step corresponding to the column.

\begin{figure}[h]
    \centering
    \subfloat[][Histograms of immune status]{\includegraphics[width=0.9\textwidth]{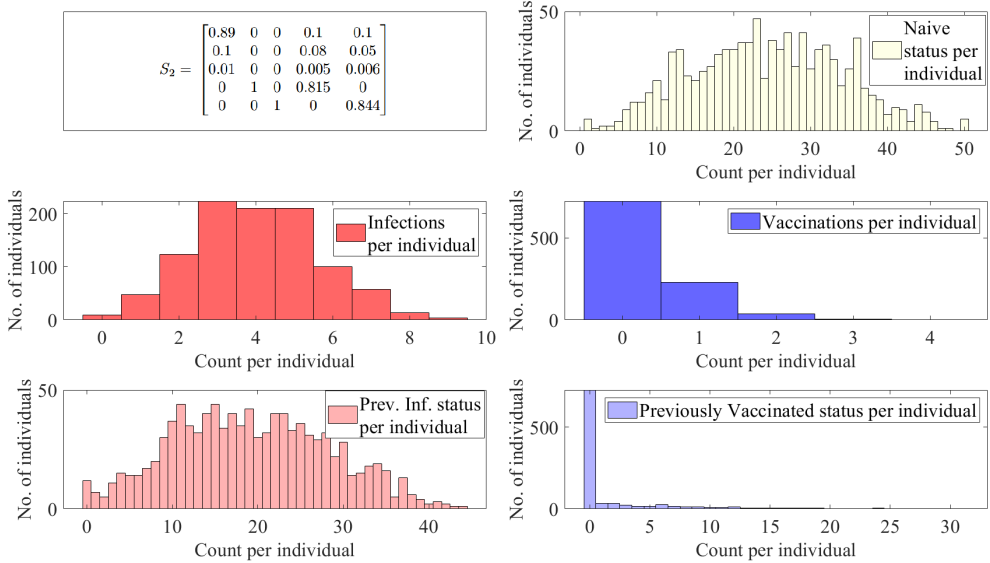} \label{fig:S2_counts}} \\
    \subfloat[][Five representative individual sequences]{ \includegraphics[width=0.9\textwidth]{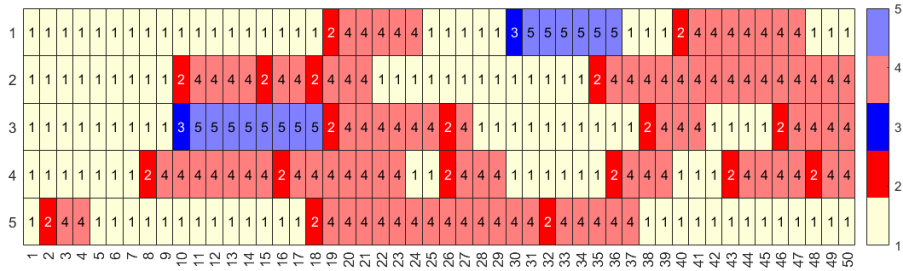} \label{fig:S2_sequences}}
    \caption{Simulations for a population with $1000$ individuals with a time-homogeneous transition matrix $S_2$, per $50$ time steps, i.e., approximately two year period. (a) Histograms of counts of total infections, vaccinations, previously infected status, previously vaccinated status, and na{\"i}ve status.
    (b) Five  representative sequences. 
    Light yellow denotes the na{\"i}ve state (1, $N$), red and pink  the newly and previously infected states (2, $\pos$ and 4, $\posp$), and blue and light blue the newly and previously vaccinated states (3, $\vac$ and 5, $\vacp$).}
\end{figure}

We increase the probability of transition to na{\"i}ve from both previously infected and previously vaccinated status to obtain $S_2$ (Figure \ref{fig:S2_counts}).
This emulates diseases with shorter-term immunity and thus a quicker return to the na{\"i}ve status; see  Figure \ref{fig:S2_counts}. As anticipated, this increases the likelihood of sequences with a higher number of immune events; see Figure  \ref{fig:S2_sequences}.

\section{Discussion}
\label{sec:disc}

In this section, we discuss a  generalization of the two-event time-inhomogeneous model, more on the general time-inhomogeneous model, and then investigate the variability of antibody response across a population. We include a comparison of the two datasets used in our validation of our methods, and finish with limitations and extensions of our work, as well as implications for immunologists.

\subsection{Two-event time-inhomogeneous model}
 
 We first note that our optimizations via MLE for the results shown in   Section \ref{sec:results_data} are robust to the choice of initial guess. Further, for the two-event immune sequence, we  only need to estimate three parameters for each event, as we take the parameters from the first event as unchanged inputs to the second event, thus improving the usability of our model.
 
Although limited in its ability to capture real world situations of an arbitrary number of infections and vaccinations, the two event model is an important stepping stone that helps us better understand the general, time-inhomogeneous setting. The general model is elegant for its ability to capture all possibilities while requiring only five states; the trade-off is the complexity of the mathematics needed to enumerate all possible immune trajectories in such a framework. As an alternative, the two event model can be generalized to increase the number of possible states at every time step. A formula for the number of states $N_s$ as a function of number of immune events $N_e$ allowed is given by
\begin{equation}
    N_s(N_e) = 4(2^{N_e} - 1) + 1.
\end{equation}
This is because each time an additional immune event is allowed, twice the number of previous-level states are created, since an individual can either become infected or vaccinated. Then, they move with probability 1 into a previous infection or previous vaccination state. The constant 1 term counts the na{\"i}ve state $N$. Allowing the number of states to increase at each time step would result in a quickly growing sequence, with 125 states necessary to model 5 possible immune events.

 In our current framework, one needs to know exactly \textit{how many} events happened and \textit{when} they happened. For the two event personal timeline model,  we speculate that one may be able to write $R_{\pos, \pos}(\bm{r}, t, \tau)$ as $R_{\pos, \pos}(\bm{r}, \bm{u}, t)$, where $\bm{u}$ is a second measured value that is known in distribution sense rather than deterministic. This could potentially be extended to a framework such that if you know your antibody level on some day, and that \textit{some} number of previous events have happened, one can predict measurements beyond this day as a sort of forecast with uncertainty ranges from possible previous events.

\subsection{General time-inhomogeneous model}

Towards constructing the conditional probability models in the general setting of the time-inhomogeneous model, we discuss an important check. We want to ensure that the number of high-probability sequences with a large number of events is vanishingly small to reflect real world situations, as a person will not realistically become reinfected or revaccinated at every point in time. Notice that with the possibility of either infection or vaccination at every time step, we obtain $2^T$ possible event sequences over $T$ time steps. In reality, we can constrain this number by quite a bit.  We note that as class $\pos$ necessarily transitions to class $\posp$ and class $\vac$ to class $\vacp$, there are at the most $ \lfloor T/2 \rfloor$ events possible before time $T$, which means there are approximately $2^{\lfloor T/2 \rfloor}$ ways for events to occur before time $T$. We recall also that time will likely be batched by several weeks, so $T = 10$ might mean 210 days if $\Delta T = 21$ days or 3 weeks, which is a reasonable time spacing. Then for this example, there would be $2^5 = 32$ possible sequences in 210 days, which is a manageable number. We can further constrain this set by calculating the probability of obtaining such a sequence as a product of the associated transition probabilities. What we need to additionally consider are the days spent between two immunological events in the previously infected or previously vaccinated state as this affects the antibody response distribution. As an illustration, consider $T = 6$. For simplicity, let us assume that $z_i \in \{\pos,\vac\}$. The 43 total trajectories are:
\begin{itemize}
    \item No immune event,  $NNNNNN$. 1 trajectory.
    \item One immune event, $N\cdots Nz_1\tilde{z}_1\cdots \tilde{z}_1$. 10 trajectories.
    \item Two immune events, $N\cdots N z_1\tilde{z}_1\cdots \tilde{z}_1z_2\tilde{z}_2\cdots \tilde{z}_2$. 24 trajectories.
    \item Three immune events, $Nz_1\tilde{z}_1z_2\tilde{z}_2z_3$. 8 trajectories.
\end{itemize}

 \subsection{Personal trajectories of interest}

 Our models provide population-level information; examining individual trajectories may provide insight into those who stay within regions of high probability versus outliers.
 This can potentially be used to predict antibody response to future infections and/or vaccinations, or suggest an individualized booster schedule, e.g., if your antibody response drops quickly, perhaps more frequent vaccine booster doses are needed. Identifying correlations in antibody trends and demographic factors may ameliorate such efforts. We leave a thorough evaluation and discussion of antibody response differences across demographic groups to \cite{ohanlon2025time}. 

 We study personal trajectories of interest for a few subjects from both datasets. To discuss individual subjects, we standardize and anonymize  the IDs by dataset (D1 or D2 for Dataset 1 or Dataset 2) and appearance number in the following figures by dataset (e.g., D1-1).  In   Figure \ref{fig:single_traj_of_interest}, we plot the single-event models for infection and vaccination with representative and outlier personal trajectories of interest.
 
 \begin{figure}[h]
    \centering
\subfloat[][Infection, Dataset 1]{\includegraphics[width=0.4\linewidth]{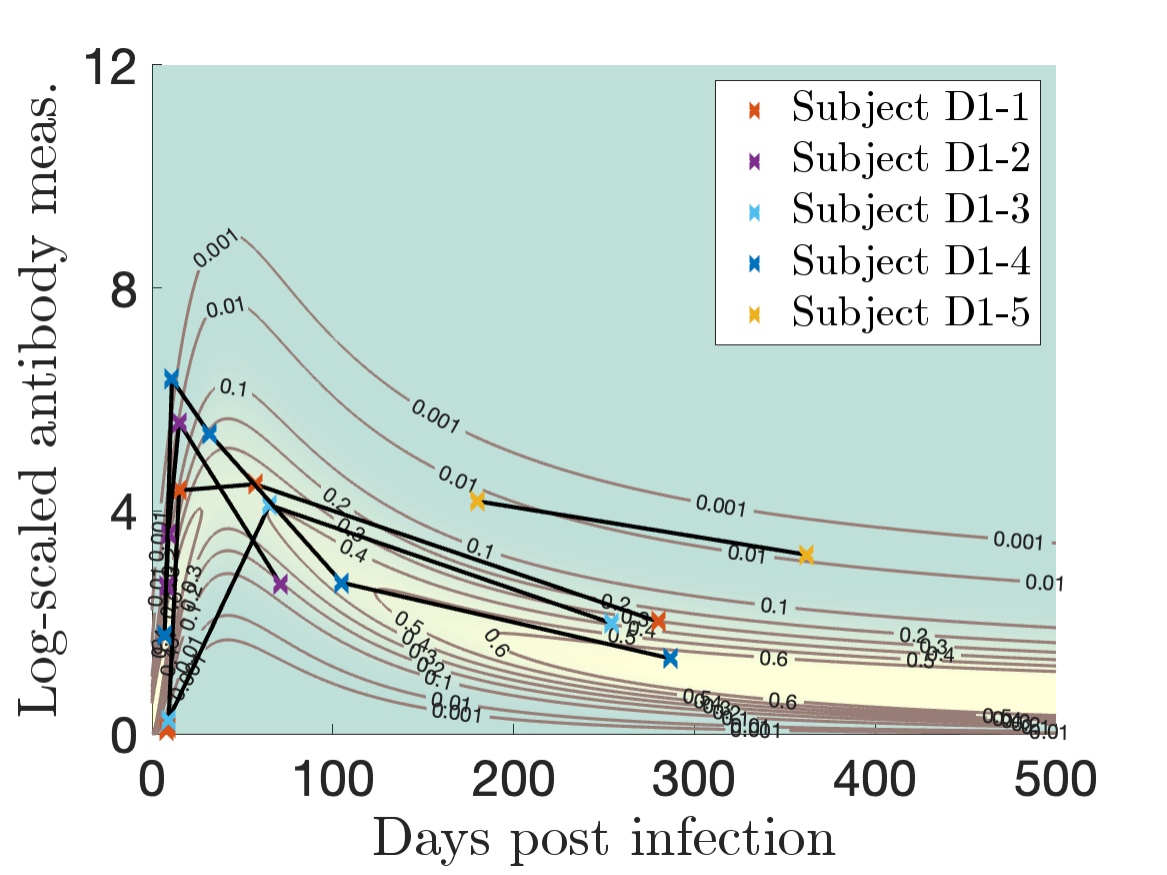}\label{fig:UVA_traj_interest_inf}}
\subfloat[][Infection, Dataset 2]{\includegraphics[width=0.4\linewidth]{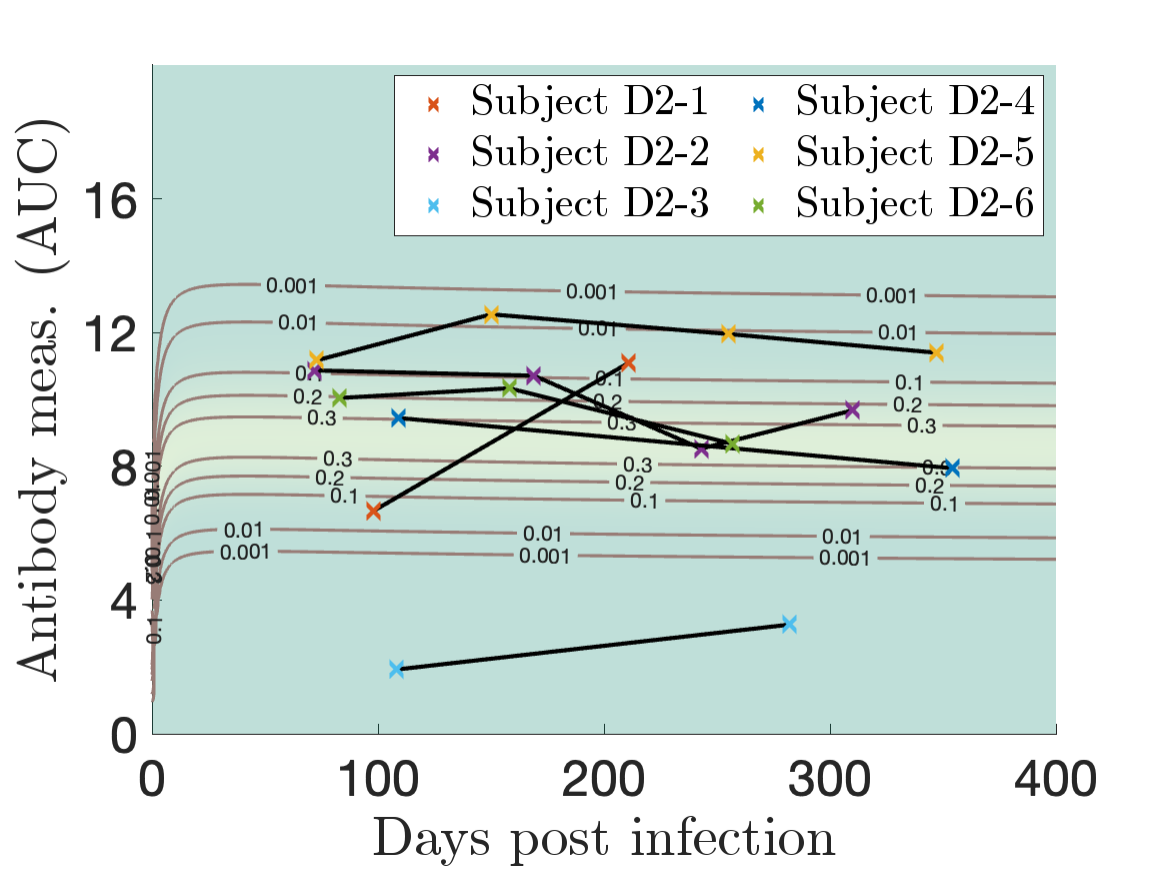}
\label{fig:UCSD_traj_interest_inf}} \\
\subfloat[][Vaccination, Dataset 1]{\includegraphics[width=0.4\linewidth]{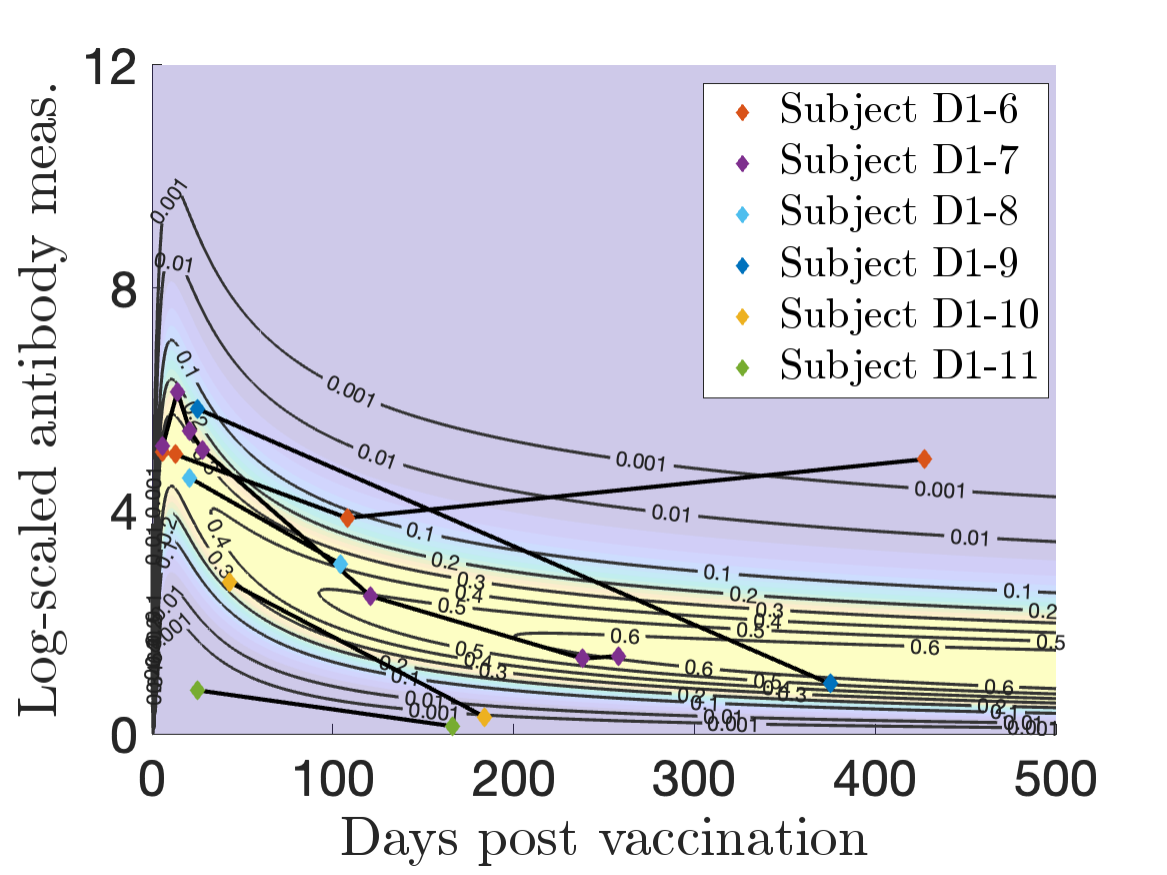}\label{fig:UVA_traj_interest_vax}} 
    \subfloat[][Vaccination, Dataset 2]{\includegraphics[width=0.4\linewidth]{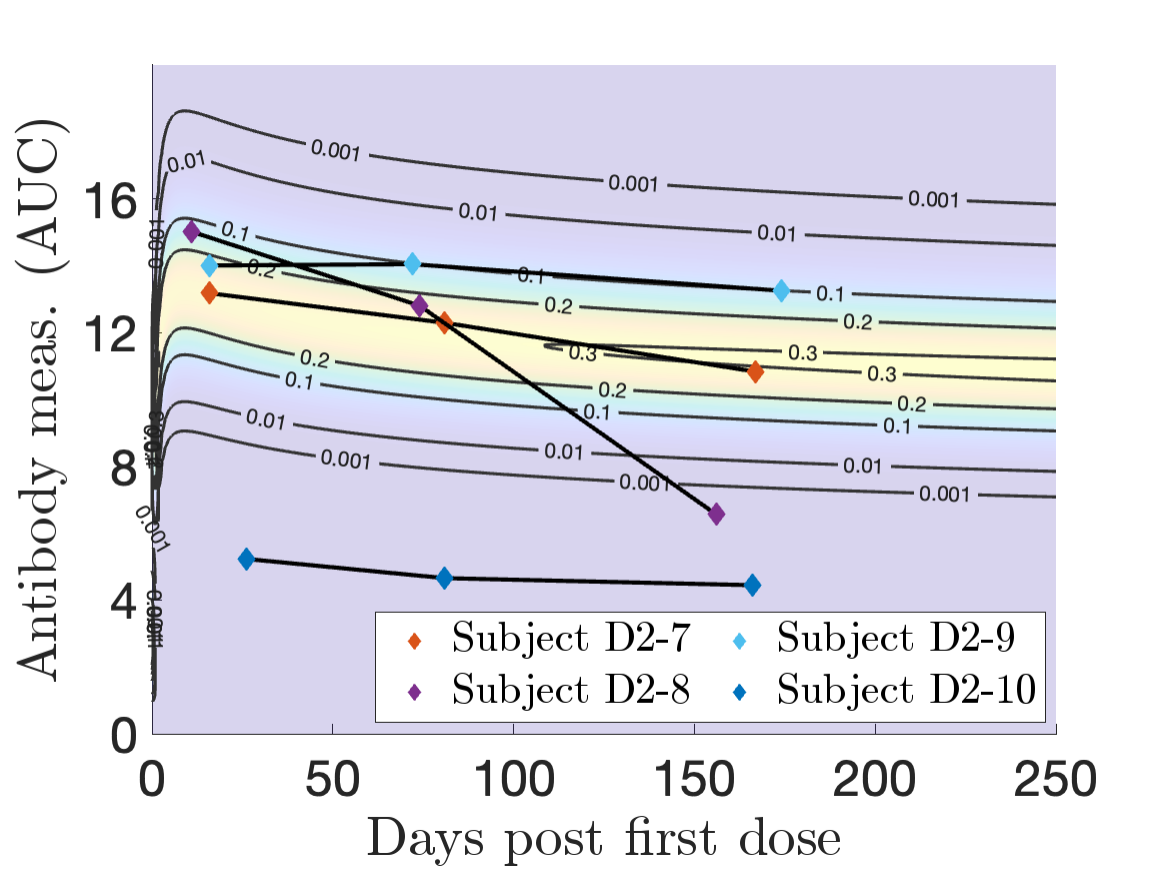}    \label{fig:USCD_traj_interest_vax}}
    \caption{Log-transformed antibody measurements from the infected (I) (a),(b) and vaccinated (V) (c),(d) populations from Dataset 1 (a),(c) and Dataset 2 (b),(d) with personal trajectories of interest with corresponding probability models. For Dataset 2, measurements with recorded monotonically decreasing OD values have been AUC-extrapolated using the next titration level.}
    \label{fig:UVA_traj_interest}
\label{fig:single_traj_of_interest}
\end{figure}

 \paragraph{Single infection.} For Dataset 1, in  Figure \ref{fig:UVA_traj_interest_inf}, we see that Subject D1-5 maintains a high response and Subjects D1-1, D1-3, and D1-4  illustrate medium responses after the initial acute phase that correspond to relatively high probability values for all involved measurements. However, Subject D1-1 shows a rapid but medium response initial increase,  Subject D1-4 has a rapid, large initial increase, and Subject D1-3 has a slower and medium response initial increase in antibody level. Subject D1-2 has a rapid, high initial response and then a quick decay to a below average antibody level. 

For Dataset 2, in  Figure \ref{fig:UCSD_traj_interest_inf}, Subject D2-5 exhibits a high response, Subject D2-3 demonstrates a low response, and Subjects D2-2, D2-6, and D2-4 illustrate medium responses that correspond to relatively high probability values for all involved measurements. We speculate that Subject D2-1 had a missed second infection because their antibody response defies the trend, increasing with days post infection.

\paragraph{Single vaccination.} Missed events are unexpected for vaccination since immunization doses are well-recorded, but there is still population variability in the durability of antibody response to a vaccine. For Dataset 1, in Figure  \ref{fig:UVA_traj_interest_vax}, we see that Subject D1-8  shows a medium response after the initial vaccination. Subjects D1-10 and D1-11 are low responders, as Subject D1-11 never mounts much of an antibody response at all, and Subject D1-10's response falls below the lowest model contour before 200 days after vaccination. Subjects D1-7 and D1-9 have relatively high initial responses, but Subject D1-7's antibody level falls much more quickly than Subject D1-9. Subject D1-6 exhibits a medium initial response, but this is sustained better than many subjects beyond 100 days, and then we speculate that they experienced a missed infection sometime before their next measurement (over 400 days after vaccination).


For Dataset 2, in  Figure \ref{fig:USCD_traj_interest_vax}, Subject D2-9 exhibits a high response, Subject D2-7 demonstrates a medium response, and Subject D2-9 illustrates a low response. Subject D2-8 shows a response that drops sharply, defying the trend of slow decay.

\begin{figure}[h]
    \centering
\includegraphics[width=1.0\linewidth]{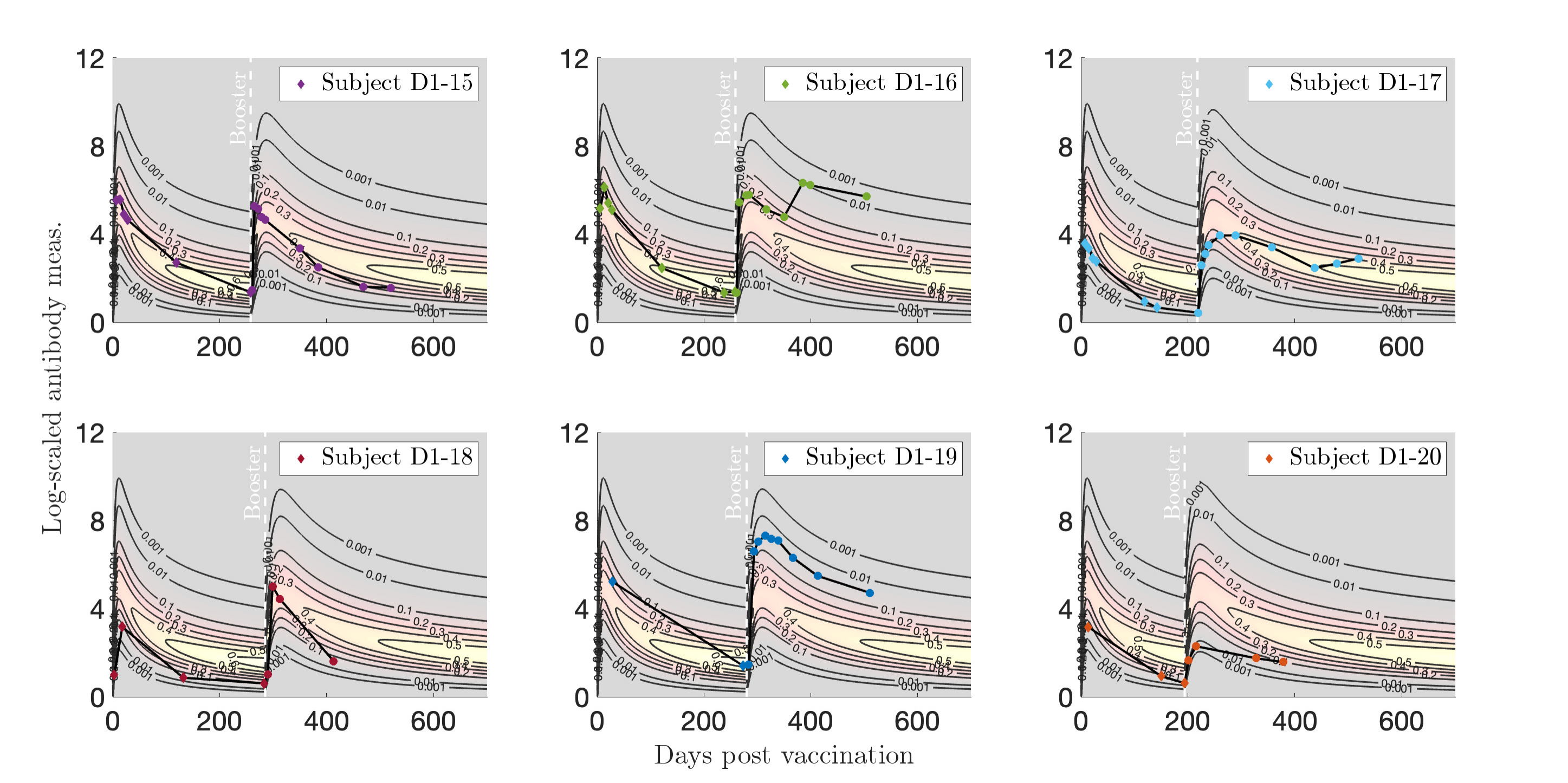}
    \caption{Log-transformed antibody measurements from the vaccinated then boosted (VV) population from Dataset 1 with personal trajectories of interest with corresponding probability models. The relative date of the booster dose is marked with a vertical, white dashed line. See this \href{https://drive.google.com/file/d/1d-M6Z4c50kTCqS0h0z5fGB3iF_t7GWhA/view?usp=sharing}{Supplementary Figure (hyperlink)} for the corresponding video and Appendix \ref{sec:supp_video_captions} for a description.}
    \label{fig:UVA_traj_interest_vax_vax}
\end{figure}

\paragraph{Two vaccinations.} In Figure  \ref{fig:UVA_traj_interest_vax_vax} we plot individualized two-event models for a booster dose after vaccination for six subjects of interest from Dataset 1. Subject D1-15 exhibits a medium response to both infection and vaccination, with representative decay patterns that are well captured by our model. This is true for Subject D1-16 for the measurements up until the booster dose, roughly 100 days after which we speculate that a breakthrough infection occurred, since the subject's antibody response spikes again. We observe that Subject D1-17 mounted a relatively low response to the first vaccine sequence, but a stronger response was induced by the booster dose. Subject D1-18 exhibits a low initial vaccine response that drops quickly, followed by a medium response to the booster that also rapidly drops. Finally, Subjects D1-20 and D1-19 show low and high antibody responses to both the initial and booster vaccinations, respectively, with a particularly high and sustained response to the booster dose for Subject D1-19. 

\begin{figure}[h]
\centering
    \includegraphics[width=0.97\linewidth]{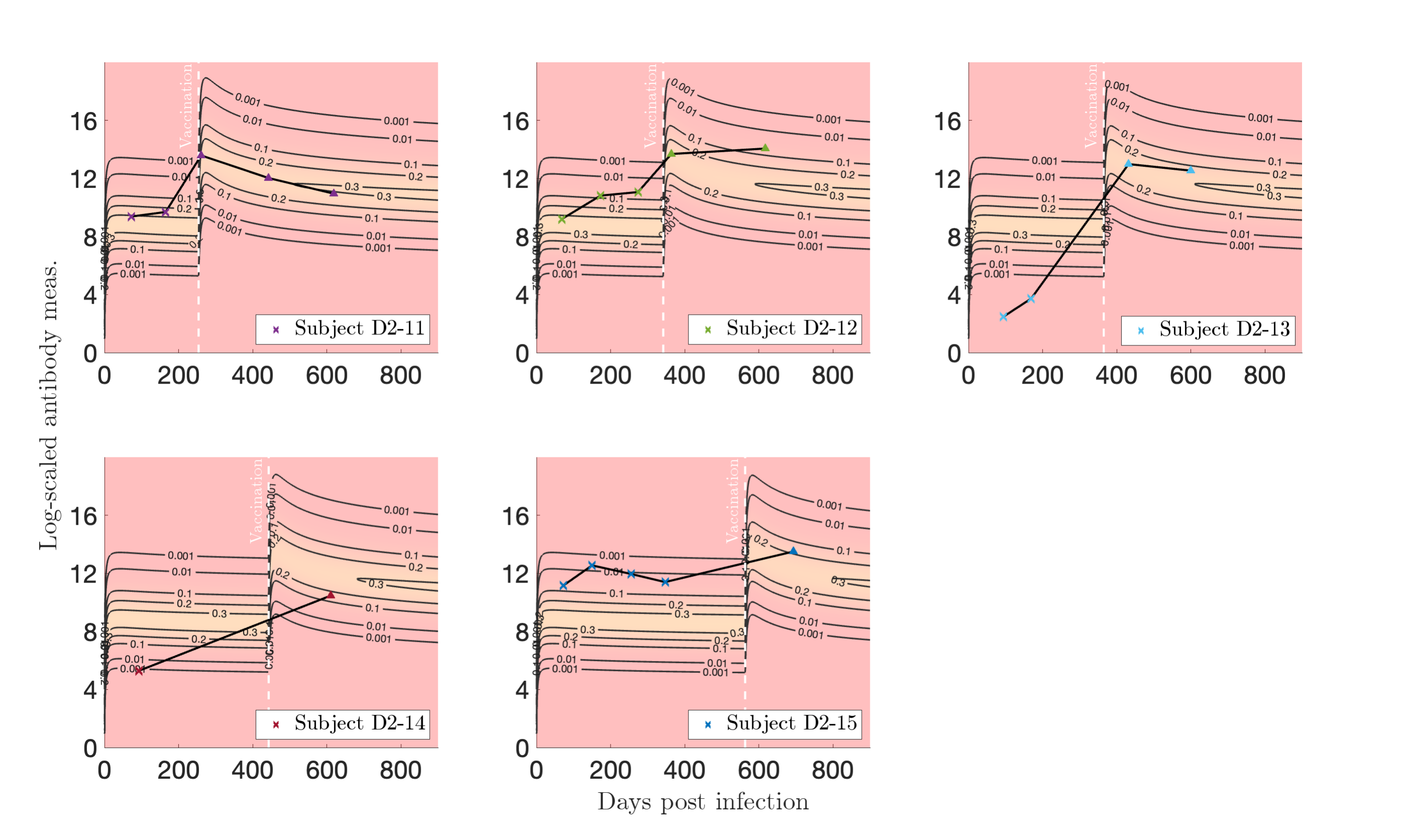}
    \caption{Log-transformed antibody measurements from the  infected, then vaccinated (VI) population from Dataset 2 with personal trajectories of interest with corresponding probability models. Measurements with recorded monotonically decreasing OD values have been AUC-extrapolated using the next titration level. See this \href{https://drive.google.com/file/d/1B2eegdxE-8SE14WdH38TpikJToesQtCJ/view?usp=sharing}{Supplementary Figure (hyperlink)} for the corresponding video and Appendix \ref{sec:supp_video_captions} for a description.}
    \label{fig:UCSD_inf_vax_traj_interest}
\end{figure}

\paragraph{Vaccination after infection.} For Dataset 2, in Figure  \ref{fig:UCSD_inf_vax_traj_interest}, we see that Subject D2-11 exhibits a medium response to both infection and vaccination, whereas a  low response to both is observed for Subject D2-14. Subjects D2-12 and D2-15 have above-average responses to both infection and vaccination, and in particular, Subject D2-12 exhibits an increasing trajectory, which is against the trend and prediction of our model. Finally, Subject D2-13 shows a low response to infection, but responds well to vaccination. A similar analysis for subjects from Dataset 1 is shown and discussed in Appendix \ref{sec:add_pers_traj_ex} (see Figure \ref{fig:UVA_traj_interest_inf_vax}).

\subsection{Comparison of Datasets 1 and 2}

Datasets 1 and 2 were recorded using different measurement systems \cite{keshavarz2022trajectory,canderan2025distinct,congrave2022twelve}; the first in IU/mL, and the second in unitless AUC. Even after titration-extrapolation for Dataset 2 (Section \ref{sec:dataset_2}), the antibody decay is more pronounced or visible in Dataset 1 than Dataset 2, as evidenced by the $k_1$ and $k_2$ values shown in  Table \ref{table:MLE_params_UVA} and Table \ref{table:MLE_params_UCSD} ($k_1$ and $k_2$ are proxies for decay speed); this is likely due to the varied measurement recording systems. For Dataset 1, we find a faster decay for infection than vaccination, but the reverse is true for Dataset 2. We find the vaccinated measurements to be more clustered across time for Dataset 2, as evidenced by the fact that 97 \% of measurements fall within the model contours, with 67 \% at a ``high'' probability level, as compared to 71 \% and 49 \% for Dataset 1. For infection: 60 \% are at a ``high'' probability level for Dataset 1, whereas only 21 \% of measurements are for Dataset 2, indicating greater variability in the second population's response to infection.

\subsection{Limitations and extensions}

Modeling is inherently subjective \cite{smith2013uncertainty}; the form of the na{\"i}ve model and shape functions for the personal response models are choices, but the influence of this issue is lessened as more sample points are used \cite{schwartz1967estimation}. As a remedy, one can select a model with minimal error on a measure of interest, such as prevalence estimates, from a proposed family of models \cite{patrone2024minimizing}. We note that immunocompromised individuals affect the modeling exercise. In the future, this could be addressed by separating the population by immune system status and correcting errors due to reporting bias or gaps.
  Uncertainty quantification could address the deterioration of accuracy and precision of collected antibody data since the start of the pandemic. Vaccination may be well-documented for a population if de-identified medical data are available, but new infections are prone to missing responses and errors due to the inexactness of days post symptom onset (DPSO) as an infection marker. Further, DPSO may often underestimate the true time since the beginning of infection.

We note that  for two or more events, our shape parameter $\alpha_n$ that describes the time-dependent nature of the personal response no longer asymptotes to $\alpha_N$ as $t \to \infty$; instead, we have
\begin{equation}
    \lim_{t \to \infty} \alpha_n(t) =  \left( \sum_{i = 1}^{n-1} \frac{\theta_i \tau_i}{1 + \phi_i \tau_i^{k_i}} \right) + \alpha_N.
\end{equation}
We expect this discrepancy to be small. In all example two-event models, the probability distribution has decreased towards similar levels to that of the na{\"i}ve distribution by the final time shown. The discrepancy is smallest for events with large relative times between them, as would be expected if one was obtaining an annual booster dose of a vaccine, for example. Future models could be designed to enforce the limiting behavior from our prior work \cite{bedekar2022prevalence,bedekar2025prevalence}.



A remaining assumption in our modeling is that the entire population starts in the na{\"i}ve state, which characterizes an emergent disease. This simplifies some of our work, including the ability to set the disease and vaccination prevalence on day zero to zero.  In the future, one could design a framework for a currently circulating disease, which could be validated using real incidence rates and antibody measurements. 

In future work, we will employ the absolute and the personal timeline models to represent the relevant probabilities for the general, time-inhomogeneous setting by enumerating all possible personal trajectories, which will likely involve a challenging application of enumerative combinatorics. Some guiding principles inform this process. Due to the biology, extreme sequences of events, such as an individual becoming newly infected every day, will be assigned very low probabilities. Further, despite the multitude of potential sequences of events leading to an antibody response and current state on a particular time step, the likelihood of infection or vaccination on the next time step depends solely on the current state information.

\subsection{Implications for immunologists}

We have created a cohesive framework for population-level multiclass antibody kinetics and transmission of or vaccination against a disease.  Our models predict the antibody response to sequences of infection and vaccination over time, and may provide useful information when considering the need for and the timing of vaccine boosters. Although we use SARS-CoV-2 as a motivating example, this approach is fully generalizable to other diseases that exhibit waning immunity, such as influenza, respiratory syncytial virus (RSV), and pertussis. In particular, the models follow biological assumptions that can be adapted or narrowed to focus on populations of interest, such as children, the elderly, or the immunocompromised. Calculated model parameters, $(\theta, k)$ could also facilitate comparison of response magnitude and durability following different immune events and in patient populations of interest.
Our methods do not make any assumptions about the time-dependent prevalences or incidence rates; instead, the data can guide us in estimating these as in previous works. An advantage of this method is that it works irrespective of policies enacted and vaccine hesitancy in the underlying population, but a disadvantage is that such prevalence estimation will require a large sample size. As assay standardization is not fully achieved, such studies should use the same data collection methods, instruments, and protocol to facilitate the comparison of measurements across large periods of time.

\section{Relevant mathematical background}
\label{sec:relevant_math_background}

\begin{figure}[h]
    \centering
    \includegraphics[width=0.9\linewidth]{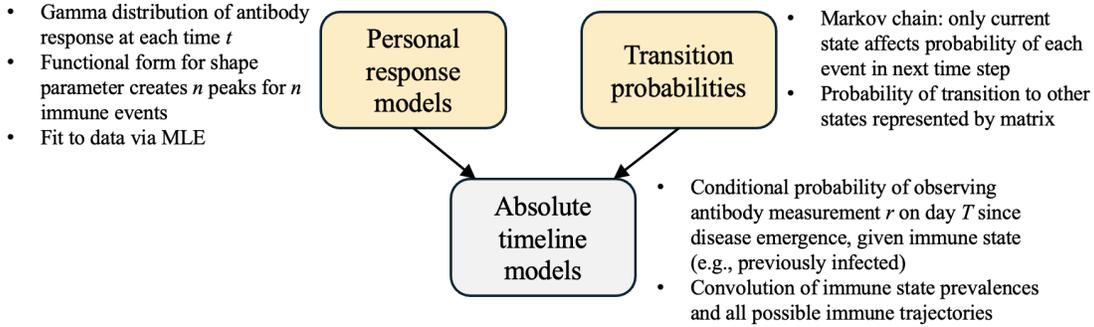}
    \caption{Schematic summarizing the main mathematical ideas in this paper.}
\label{fig:appendix_schematic}
\end{figure}

 The central thesis of our work is that an antibody measurement's interpretation relies on when the sample was obtained, the prior individual immune events, and the prevalence of events in the population at the time. To that end, this work employs probabilistic modeling to analyze personal antibody kinetics post-immune events such as infection and vaccination. In parallel, we apply time-inhomogeneous Markov chains to study population-level incidence rates and prevalence. Figure \ref{fig:appendix_schematic} shows a schematic that summarizes how the main mathematical ideas of our work fit together.  An important accomplishment of this work is the simultaneous tracking of antibody response and immune state across multiple immune events. 
 
\paragraph{Personal response models (Figure \ref{fig:appendix_schematic}, left).} We first construct probabilistic models of antibody kinetics; such models account for the variation of antibody measurements across individuals. We use the gamma probability density; refer to the na{\"i}ve model in Figure \ref{fig:UVA_models_k_traj}, for instance. Along the $x$-axis, we have plotted log-scaled antibody levels, while the $y$-axis denotes their relative abundance in the population. 

 When the disease persists in the population, individuals face multiple immune events, such as reinfection, booster vaccination, or combinations like vaccination followed by infection and vice versa.  We model the antibody kinetics of such immune events sequences by considering factors such as the time between immune events.  The addition of a new dimension--time since immune event--necessitates a new model depiction.
For the models that characterize responses to infection, vaccination, or repeated exposures (e.g., the rest of Figure \ref{fig:UVA_models_single}), each vertical slice in time $t$ represents the probability distribution  across antibody levels. The $x$-axis denotes the days since the immune event, the $y$-axis denotes the antibody measurement, and the contour lines denote the relative abundance in the population. 
We strongly encourage readers to watch the videos (described in Appendix \ref{sec:supp_video_captions}) to visualize of our results. 
The models are fit to real immune event antibody data  via Maximum Likelihood Estimation (MLE), a statistical technique that estimates the parameters associated with a probability distribution that maximizes the likelihood of observing the given data.

\paragraph{Transition probabilities (Figure \ref{fig:appendix_schematic}, right).}   At the population level, we model how infections and vaccinations spread and accumulate using a time-inhomogeneous Markov chain (see \eqref{eq:general_trans_prob}, \eqref{eq:trans_mat_1step}, \eqref{eq:H_transition_general}). We specifically study emergent infectious diseases, 
meaning that immunity emerges \textit{de novo}, unconfounded by earlier or undocumented immune events.
Over time, a fraction of the population moves to the acutely/newly infected (\pos) state with some time-dependent incidence rate. This fraction then becomes previously infected (\posp), where an individual possesses time-varying antibody protection that peaks and then wanes. Once a vaccine is developed, analogous states exist for this immune event (\vac, \vacp).    
In particular, \eqref{eq:trans_mat_1step} is a matrix whose entries give the probability of transition from one state to another at a given time $T$.  Markov chains are a reasonable approach  to track how individuals transition between immune states because the probability of an immune event in the upcoming time step depends on the current immune state. Time-inhomogeneity refers to the fact that the incidence rates and prevalences of these immune events change over time.

\paragraph{Absolute timeline models (Figure \ref{fig:appendix_schematic}, bottom).}  We combine the personal kinetics and population level changes to model the conditional probability of an antibody measurement. Antibody measurements obtained at a given time from a random sample of the population can be explained as a convolution of the effects of personal antibody kinetics and the population-level prevalences. 
This defines the probability of obtaining an antibody measurement at a particular time given that the individual faced a certain sequence of immune events so far (see Section \ref{sec:cond_prob_two_event} and Section \ref{subsec:cond_prob_time_inhomogeneous}).


\subsection{Further reading}

\begin{itemize}
    \item \textit{An Introduction to Mathematical Statistics and Its Applications}, Larsen and Marx (conditional probability, the gamma distribution, maximum likelihood estimation)
    \item \textit{Finite Markov Chains and Algorithmic Applications}, H{\"a}ggstr{\"o}m (Markov chains, including time-inhomogeneous Markov chains)

    \item  \textit{Plotkin’s Vaccines},
  Orenstein, Offit, Edwards, \& Plotkin (vaccine history, immunology, technology)

\item  \textit{SARS-CoV-2-infection- and vaccine-induced antibody responses are long lasting with an initial waning phase followed by a stabilization phase}, Srivastava et al.
Immunity, 57 (2024) pp. 587 - 599 (characteristics of antibody kinetics, hybrid immunity of infection and vaccination)
\end{itemize}

\section*{Declarations}

\paragraph{Authorship and contributorship.} All authors have made substantial intellectual contributions to the study conception, execution, and design of the work. All authors have read, reviewed, edited, and approved the final manuscript.  In addition, the following contributions occurred: 
Rayanne A. Luke: Conceptualization, Methodology, Software, Formal analysis, Writing - original draft preparation, Visualization, Funding acquisition.
Prajakta Bedekar: Conceptualization, Methodology, Software, Formal analysis, Writing - original draft preparation, Visualization.
Lyndsey M. Muehling: Investigation, Data curation.
Glenda Canderan: Investigation, Data curation.
Yesun Lee: Investigation, Data curation.
Wesley Cheng: Investigation, Data curation.
Judith A. Woodfolk: Investigation, Resources, Funding acquisition.
Jeffrey M. Wilson: Investigation, Resources, Funding acquisition.
Pia S. Pannaraj: Conceptualization, Investigation, Resources, Funding acquisition.
Anthony J. Kearsley: Conceptualization, Methodology, Supervision, Funding acquisition.

\paragraph{Conflicts of interest.} The authors declare there are no conflicts of interest.

\paragraph{Acknowledgements.} This work is a contribution of the National Institute of Standards
and Technology, USA and is not subject to copyright in the United States. Use of data in this manuscript has been approved by the NIST
Research Protections Office (study number ITL-2025-0431).  The authors thank Bradley Alpert, Melinda Kleczynski, and Amanda Pertzborn for useful feedback during preparation of this manuscript.

\paragraph{Data \& code availability.} The authors plan to make code available on GitHub once our paper is accepted for publication. Original data are from Keshavarz et al. \cite{keshavarz2022trajectory}, Canderan et al. \cite{canderan2025distinct}, and Congrave-Wilson et al. \cite{congrave2022twelve}; see their data availability statements for how to request or access relevant data.

\paragraph{Funding.} R.A.L. was funded by 4-VA, a collaborative partnership for advancing the Commonwealth of Virginia.  P.B. was funded through the NIST PREP, USA grant 70NANB18H162. J.A.W. was funded through NIH R21 AI 160334 and R56 AI 178669. J.M.W. was funded through the University of Virginia Manning COVID-19 Research Fund. P.S.P. was funded through NIH NIAID R01AI173194. The aforementioned funders had no role in study design, data analysis, decision to publish, or manuscript preparation.

\bibliographystyle{unsrt}

\bibliography{bibliography.bib}

\begin{thebibliography}{10}

\bibitem{srivastava2024sars}
Komal Srivastava, Juan~Manuel Carre{\~n}o, Charles Gleason, Brian Monahan,
  Gagandeep Singh, Anass Abbad, Johnstone Tcheou, Ariel Raskin, Giulio Kleiner,
  Harm van Bakel, et~al.
\newblock {SARS-CoV-2}-infection-and vaccine-induced antibody responses are
  long lasting with an initial waning phase followed by a stabilization phase.
\newblock {\em Immunity}, 57(3):587--599, 2024.

\bibitem{diep2023successive}
Anh~Nguyet Diep, Joey Schyns, Claire Gourzon{\`e}s, Emeline Goffin, Iraklis
  Papadopoulos, S~Moges, Fr{\'e}d{\'e}ric Minner, O~Ek, Germain Bonhomme,
  Marine Paridans, et~al.
\newblock How do successive vaccinations and {SARS-CoV}-2 infections impact
  humoral immunity dynamics: An 18-month longitudinal study.
\newblock {\em J Infect}, 2023.

\bibitem{guo2023durability}
Li~Guo, Qiao Zhang, Xiaoying Gu, Lili Ren, Tingxuan Huang, Yanan Li, Hui Zhang,
  Ying Liu, Jingchuan Zhong, Xinming Wang, et~al.
\newblock Durability and cross-reactive immune memory to {SARS-CoV}-2 in
  individuals 2 years after recovery from {COVID}-19: a longitudinal cohort
  study.
\newblock {\em Lancet Microbe}, 2023.

\bibitem{liu2023persistence}
Xinxue Liu, Alasdair~PS Munro, Annie Wright, Shuo Feng, Leila Janani,
  Parvinder~K Aley, Gavin Babbage, Jonathan Baker, David Baxter, Tanveer Bawa,
  et~al.
\newblock Persistence of immune responses after heterologous and homologous
  third {COVID}-19 vaccine dose schedules in the {UK}: eight-month analyses of
  the {COV-BOOST} trial.
\newblock {\em J Infect}, 87(1):18--26, 2023.

\bibitem{caini2020meta}
Saverio Caini, Federica Bellerba, Federica Corso, Ang{\'e}lica
  D{\'\i}az-Basabe, Gioacchino Natoli, John Paget, Federica Facciotti,
  Simone~Pietro De~Angelis, Sara Raimondi, Domenico Palli, et~al.
\newblock Meta-analysis of diagnostic performance of serological tests for
  {SARS-CoV-2} antibodies up to 25 {A}pril 2020 and public health implications.
\newblock {\em Eurosurveillance}, 25(23):2000980, 2020.

\bibitem{peeling2020serology}
Rosanna~W Peeling, Catherine~J Wedderburn, Patricia~J Garcia, Debrah Boeras,
  Noah Fongwen, John Nkengasong, Amadou Sall, Amilcar Tanuri, and David~L
  Heymann.
\newblock Serology testing in the {COVID}-19 pandemic response.
\newblock {\em Lancet Infect Dis}, 20(9):e245--e249, 2020.

\bibitem{osborne2000ten}
Kate Osborne, Nigel Gay, Louise Hesketh, Peter Morgan-Capner, and Elizabeth
  Miller.
\newblock Ten years of serological surveillance in {E}ngland and {W}ales:
  methods, results, implications and action.
\newblock {\em Int J Epidemiol}, 29(2):362--368, 2000.

\bibitem{pollan2020prevalence}
Marina Poll{\'a}n, Beatriz P{\'e}rez-G{\'o}mez, Roberto Pastor-Barriuso,
  Jes{\'u}s Oteo, Miguel~A Hern{\'a}n, Mayte P{\'e}rez-Olmeda, Jose~L
  Sanmart{\'\i}n, Aurora Fern{\'a}ndez-Garc{\'\i}a, Israel Cruz,
  Nerea~Fern{\'a}ndez de~Larrea, et~al.
\newblock Prevalence of {SARS-CoV}-2 in {S}pain ({ENE-COVID}): a nationwide,
  population-based seroepidemiological study.
\newblock {\em Lancet}, 396(10250):535--544, 2020.

\bibitem{bajema2021estimated}
Kristina~L Bajema, Ryan~E Wiegand, Kendra Cuffe, Sadhna~V Patel, Ronaldo
  Iachan, Travis Lim, Adam Lee, Davia Moyse, Fiona~P Havers, Lee Harding,
  et~al.
\newblock Estimated {SARS-CoV}-2 seroprevalence in the {US} as of {S}eptember
  2020.
\newblock {\em JAMA Intern Med}, 181(4):450--460, 2021.

\bibitem{hoze2025rsero}
Nathana{\"e}l Hoz{\'e}, Margarita Pons-Salort, C~Jessica~E Metcalf, Michael
  White, Henrik Salje, and Simon Cauchemez.
\newblock R{S}ero: A user-friendly {R} package to reconstruct pathogen
  circulation history from seroprevalence studies.
\newblock {\em PLoS Comput Biol}, 21(2):e1012777, 2025.

\bibitem{murphy2024understanding}
Quiyana~M Murphy, George~K Lewis, Mohammad~M Sajadi, Jonathan~E Forde, and
  Stanca~M Ciupe.
\newblock Understanding antibody magnitude and durability following vaccination
  against {SARS-CoV-2}.
\newblock {\em Mathematical Biosciences}, 376:109274, 2024.

\bibitem{d2020assessment}
Marco D'Arienzo and Angela Coniglio.
\newblock Assessment of the {SARS-CoV}-2 basic reproduction number, {R}0, based
  on the early phase of {COVID}-19 outbreak in {I}taly.
\newblock {\em Biosaf Health}, 2(2):57--59, 2020.

\bibitem{mcmahon2020reinfection}
Andrew McMahon, Nicole~C Robb, et~al.
\newblock Reinfection with {SARS-CoV}-2: Discrete {SIR} (susceptible, infected,
  recovered) modeling using empirical infection data.
\newblock {\em JMIR Public Health Surveill}, 6(4):e21168, 2020.

\bibitem{quick2021regression}
Corbin Quick, Rounak Dey, and Xihong Lin.
\newblock Regression models for understanding {COVID-19} epidemic dynamics with
  incomplete data.
\newblock {\em Journal of the American Statistical Association},
  116(536):1561--1577, 2021.

\bibitem{roberto2021sars}
Charles Roberto~Telles, Henrique Lopes, and Diogo Franco.
\newblock {SARS-COV-2}: {SIR} model limitations and predictive constraints.
\newblock {\em Symmetry}, 13(4):676, 2021.

\bibitem{dick2021covid}
David~W Dick, Lauren Childs, Zhilan Feng, Jing Li, Gergely R{\"o}st, David~L
  Buckeridge, Nick~H Ogden, and Jane~M Heffernan.
\newblock {COVID-19} seroprevalence in canada modelling waning and boosting
  {COVID}-19 immunity in {C}anada a {C}anadian immunization research network
  study.
\newblock {\em Vaccines}, 10(1):17, 2021.

\bibitem{stanoev2014modeling}
Angel Stanoev, Daniel Trpevski, and Ljupco Kocarev.
\newblock Modeling the spread of multiple concurrent contagions on networks.
\newblock {\em PloS One}, 9(6):e95669, 2014.

\bibitem{schuh2024mathematical}
Lea Schuh, Peter~V Markov, Vladimir~M Veliov, and Nikolaos~I Stilianakis.
\newblock A mathematical model for the within-host (re) infection dynamics of
  {SARS-CoV-2}.
\newblock {\em Mathematical Biosciences}, 371:109178, 2024.

\bibitem{hay2019characterising}
James~A Hay, Karen Laurie, Michael White, and Steven Riley.
\newblock Characterising antibody kinetics from multiple influenza infection
  and vaccination events in ferrets.
\newblock {\em PLoS Comput Biol}, 15(8):e1007294, 2019.

\bibitem{bedekar2022prevalence}
Prajakta Bedekar, Anthony~J Kearsley, and Paul~N Patrone.
\newblock Prevalence estimation and optimal classification methods to account
  for time dependence in antibody levels.
\newblock {\em J Theor Biol}, page 111375, 2022.

\bibitem{bedekar2025prevalence}
Prajakta Bedekar, Rayanne~A Luke, and Anthony~J Kearsley.
\newblock Prevalence estimation methods for time-dependent antibody kinetics of
  infected and vaccinated individuals: A {M}arkov chain approach.
\newblock {\em Bull Math Biol}, 87(2):1--33, 2025.

\bibitem{keshavarz2022trajectory}
Behnam Keshavarz, Nathan~E Richards, Lisa~J Workman, Jaimin Patel, Lyndsey~M
  Muehling, Glenda Canderan, Deborah~D Murphy, Savannah~G Brovero, Samuel~M
  Ailsworth, Will~H Eschenbacher, et~al.
\newblock Trajectory of {IgG} to {SARS-CoV}-2 after vaccination with {BNT}162b2
  or m{RNA}-1273 in an employee cohort and comparison with natural infection.
\newblock {\em Front Immunol}, 13:850987, 2022.

\bibitem{canderan2025distinct}
Glenda Canderan, Lyndsey~M Muehling, Alexandra Kadl, Shay Ladd, Catherine
  Bonham, Claire~E Cross, Sierra~M Lima, Xihui Yin, Jeffrey~M Sturek, Jeffrey~M
  Wilson, et~al.
\newblock Distinct type 1 immune networks underlie the severity of restrictive
  lung disease after {COVID}-19.
\newblock {\em Nat Immunol}, pages 1--12, 2025.

\bibitem{congrave2022twelve}
Zion Congrave-Wilson, Wesley~A Cheng, Yesun Lee, Stephanie Perez, Lauren
  Turner, Carolyn~Jennifer Marentes~Ruiz, Shirley Mendieta, Adam Skura, Jaycee
  Jumarang, Jennifer Del~Valle, et~al.
\newblock Twelve-month longitudinal serology in {SARS-CoV}-2 na{\"\i}ve and
  experienced vaccine recipients and unvaccinated {COVID}-19-infected
  individuals.
\newblock {\em Vaccines}, 10(5):813, 2022.

\bibitem{bouter2023textbook}
Lex Bouter, Maurice Zeegers, and Tianjing Li.
\newblock {\em Textbook of epidemiology}.
\newblock John Wiley \& Sons, 2023.

\bibitem{patrone2021classification}
Paul~N Patrone and Anthony~J Kearsley.
\newblock Classification under uncertainty: data analysis for diagnostic
  antibody testing.
\newblock {\em Math Med Biol}, 38(3):396--416, 2021.

\bibitem{luke2023optimal}
Rayanne~A Luke, Anthony~J Kearsley, and Paul~N Patrone.
\newblock Optimal classification and generalized prevalence estimates for
  diagnostic settings with more than two classes.
\newblock {\em Math Biosci}, 358:108982, 2023.

\bibitem{ohanlon2025time}
James O’Hanlon, Kaitlyn Sullivan, Lyndsey~M Muehling, Glenda Canderan,
  Jeffrey~M Wilson, Judith~A Woodfolk, and Rayanne~A Luke.
\newblock A probabilistic modeling analysis of the longitudinal immune response
  to infection and vaccination across demographic groups and health outcomes.
\newblock {\em In preparation}, 2025+.

\bibitem{smith2013uncertainty}
Ralph~C Smith.
\newblock {\em Uncertainty quantification: theory, implementation, and
  applications}, volume~12.
\newblock S{IAM}, 2013.

\bibitem{schwartz1967estimation}
Stuart~C Schwartz.
\newblock Estimation of probability density by an orthogonal series.
\newblock {\em Ann Math Stat}, pages 1261--1265, 1967.

\bibitem{patrone2024minimizing}
Paul~N Patrone and Anthony~J Kearsley.
\newblock Minimizing uncertainty in prevalence estimates.
\newblock {\em Stat Prob Lett}, 205:109946, 2024.

\end{thebibliography}

\newpage


\appendix

\setcounter{equation}{0}
\setcounter{figure}{0}
\setcounter{table}{0}
\renewcommand{\theequation}{A\arabic{equation}}
\renewcommand{\thefigure}{A\arabic{figure}}
\renewcommand{\thetable}{A\arabic{table}}

\section{Additional time-inhomogeneous two event model details}
\label{sec:appendix_model}

To better illustrate the time-inhomogeneous two event transition matrix $S$ given by \eqref{eq:two_event_matrix}, we explicitly calculate the probability of a person residing in a particular class after five time steps using the product of transition matrices:
\begin{equation}
\begin{split}
& S(4)S(3)S(2) S(2) S(1) S(0) \bm{e}_1 \\
& = \begin{bmatrix}
s_N(4) s_N(3) s_N(2) s_N(1) s_N(0) \\
s_{\pos N}(4) s_N(3) s_N(2) s_N(1) s_N(0) \\
s_{\vac N}(4) s_N(3) s_N(2) s_N(1) s_N(0) \\
s_{\pos N}(3) s_N(2) s_N(1) s_N(0) + s_{\posp}(4) \left\{ s_{\pos N}(2) s_N(1) s_N(0) + s_{\posp}(3) [s_{\pos N}(1) s_N(0) + s_{\posp}(2) s_{\pos N}(0)] \right\} \\
s_{\vac N}(3) s_N(2) s_N(1) s_N(0) + s_{\vacp}(4) \left\{ s_{\vac N}(2) s_N(1) s_N(0) + s_{\vacp}(3) [s_{\vac N}(1) s_N(0) + s_{\vacp}(2) s_{\vac N}(0)] \right\} \\
s_{\pos \posp}(4) \left\{ s_{\pos N}(2) s_N(1) s_N(0) + s_{\posp}(3) [ s_{\pos N} (1) s_N(0) + s_{\posp}(2) s_{\pos N}(0)] \right\} \\
s_{\vac \posp}(4) \left\{ s_{\pos N}(2) s_N(1) s_N(0) + s_{\posp}(3) [ s_{\pos N} (1) s_N(0) + s_{\posp}(2) s_{\pos N}(0)] \right\} \\
s_{\pos \vacp}(4) \left\{ s_{\vac N}(2) s_N(1) s_N(0) + s_{\vacp}(3) [ s_{\vac N} (1) s_N(0) + s_{\vacp}(2) s_{\vac N}(0)] \right\} \\
s_{\vac \vacp}(4) \left\{ s_{\vac N}(2) s_N(1) s_N(0) + s_{\vacp}(3) [ s_{\vac N} (1) s_N(0) + s_{\vacp}(2) s_{\vac N}(0)] \right\} \\
s_{\pos \posp}(3) [ s_{\pos N}(1) s_N(0)  + s_{\posp}(2) s_{\pos N}(0) ] + s_{\pos \posp}(2) s_{\pos N}(0) \\
s_{\vac \posp}(3) [ s_{\pos N}(1) s_N(0)  + s_{\posp}(2) s_{\pos N}(0) ] + s_{\vac \posp}(2) s_{\pos N}(0)\\
s_{\pos \vacp}(3) [ s_{\vac N}(1) s_N(0)  + s_{\vacp}(2) s_{\vac N}(0) ] + s_{\pos \vacp}(2) s_{\vac N}(0) \\
s_{\vac \vacp}(3) [ s_{\vac N}(1) s_N(0)  + s_{\vacp}(2) s_{\vac N}(0) ] + s_{\vac \vacp}(2) s_{\vac N}(0)
\end{bmatrix}.
\end{split}
\label{eq:5_steps_example}
\end{equation}
We interpret a few entries of \eqref{eq:5_steps_example} as follows. The first entry, Prob$(X_4 = N)$, computes the probability that one stays na{\"i}ve through time step 5; the second entry, Prob$(X_4 = \pos)$, gives the probability that one did become infected on time step 5 but did not become infected on time steps 1, 2, 3, or 4. The fourth entry, Prob$(X_4 = \posp)$, is the probability that one did not become infected on time steps 1, 2, or 3, but did become infected on time step 4 and is considered previously infected on time step 5 plus the probability that one became infected on a previous time step (computed by summing  the individual event probabilities together since they are disjoint), was then considered previously infected, and remained previously infected on subsequent time steps including time step 5. The sixth entry, Prob$(X_4 = \pos \posp)$, is the probability that one became infected on time steps 1, 2, or 3, then remained previously infected, then becomes newly infected on the fifth time step (indicated by the term $s_{\pos \posp}(4)$). The quantity in curly braces appears in the fourth entry, too, denoting infection before time step 4. The tenth entry, Prob$(X_4 = \posp \posp)$, is the probability that one became infected on time step 1 or 2 and becomes newly reinfected on time step 4 and is considered previously reinfected on time step 5, plus the probability that one became infected on time step 1, reinfected at the next possible opportunity (time step 3), and then is considered previously infected thereafter. We confirm that all possible trajectories to $N, \pos, \posp, \pos \posp,$ and $\posp \posp$ on time step 5 are considered.
Parallel interpretations hold for quantities involving $\vac$,$\vacp$, and cross events of infection after vaccination or vice versa.

Recalling from \eqref{eq:H_transition} that multi-step transitions can be represented by repeated multiplications of the matrix $S$, we generalize the above concepts to an arbitrary number of time steps $T+1$:
\begin{equation}
\begin{bmatrix}
\text{Prob}(X_T = N) \\
\text{Prob}(X_T = \pos) \\
\text{Prob}(X_T = \vac) \\
\text{Prob}(X_T = \posp) \\
\text{Prob}(X_T = \vacp) \\
\text{Prob}(X_T = \pos \posp) \\ \text{Prob}(X_T = \vac \posp) \\
\text{Prob}(X_T = \pos \vacp) \\
\text{Prob}(X_T = \vac \vacp) \\
\text{Prob}(X_T = \posp \posp) \\ \text{Prob}(X_T = \vacp \posp) \\
\text{Prob}(X_T = \posp \vacp) \\
\text{Prob}(X_T = \vacp \vacp)
\end{bmatrix} = H_T \bm{e}_1.
\label{eq:state_vec}
\end{equation}

By using \eqref{eq:state_vec}, the state probabilities are given explicitly by
\begin{subequations}
\begin{align}
\text{Prob}(X_T = N) & =     \prod_{t = 0}^T s_N(t),  \\
\text{Prob}(X_T = \pos) & = s_{\pos N}(T) \prod_{t = 0}^{T-1} s_N(t), \\
\text{Prob}(X_T = \vac) & = \displaystyle s_{\vac N}(T) \prod_{t = 0}^{T-1} s_N(t), \\
\text{Prob}(X_T = \posp) & = \sum_{t = 0}^{T-1}  \left( \prod_{\sigma = t+2}^T s_{\posp}(\sigma) \right) s_{\pos N}(t)  \prod_{\ell = 0}^{t-1} s_N(\ell) , \\
\text{Prob}(X_T = \vacp) & = \sum_{t = 0}^{T-1}  \left( \prod_{\sigma = t+2}^T s_{\vacp}(\sigma)  \right) s_{\vac N}(t)   \prod_{\ell = 0}^{t-1} s_N(\ell) , \\
\text{Prob}(X_T = \pos \posp) & = s_{\pos \posp}(T)  \sum_{t = 0}^{T-2} \left(\prod_{\sigma = t+2}^{T-1} s_{\posp}(\sigma) \right) s_{\pos N}(t)  \prod_{\ell = 0}^{t-1} s_N(\ell) , \\
\text{Prob}(X_T = \vac \posp) & = s_{\vac \posp}(T)  \sum_{t = 0}^{T-2}  \left( \prod_{\sigma = t+2}^{T-1} s_{\posp}(\sigma) \right) s_{\pos N}(t) \prod_{\ell = 0}^{t-1} s_N(\ell) , \\
\text{Prob}(X_T = \pos \vacp) & = s_{\pos \vacp}(T)  \sum_{t = 0}^{T-2} \left( \prod_{\sigma = t+2}^{T-1} s_{\vacp}(\sigma) \right) s_{\vac N}(t)  \prod_{\ell = 0}^{t-1} s_N(\ell) , \\
\text{Prob}(X_T = \vac \vacp) & = s_{\vac \vacp}(T)  \sum_{t = 0}^{T-2}   \left( \prod_{\sigma = t+2}^{T-1} s_{\vacp}(\sigma) \right) s_{\vac N}(t) \prod_{\ell = 0}^{t-1} s_N(\ell) , \\
 \text{Prob}(X_T = \posp \posp) & = \sum_{t = 2}^{T-1} s_{\pos \posp}(t) \sum_{i = 0}^{t-2} \left( \prod_{\sigma = i+2}^{t-1} s_{\posp}(\sigma) \right) s_{\pos N}(i)  \prod_{\ell = 0}^{i-1} s_N(\ell), \\
  \text{Prob}(X_T = \vacp \posp) & = \sum_{t = 2}^{T-1} s_{\vac \posp}(t) \sum_{i = 0}^{t-2} \left( \prod_{\sigma = i+2}^{t-1} s_{\posp}(\sigma) \right) s_{\pos N}(i)  \prod_{\ell = 0}^{i-1} s_N(\ell), \\
   \text{Prob}(X_T = \posp \vacp) & = \sum_{t = 2}^{T-1} s_{\pos \vacp}(t) \sum_{i = 0}^{t-2} \left( \prod_{\sigma = i+2}^{t-1} s_{\vacp}(\sigma) \right) s_{\vac N}(i)  \prod_{\ell = 0}^{i-1} s_N(\ell),\\
    \text{Prob}(X_T = \vacp \vacp) & = \sum_{t = 2}^{T-1} s_{\vac \vacp}(t) \sum_{i = 0}^{t-2} \left( \prod_{\sigma = i+2}^{t-1} s_{\vacp}(\sigma) \right) s_{\vac N}(i)  \prod_{\ell = 0}^{i-1} s_N(\ell),
    \end{align}
    \label{eq:state_prob_full}
\end{subequations}
where we adopt the convention that 
if the top index of a product is greater than the bottom index, the result is the empty product, whose value is 1.

\section{Additional population simulation example}
\label{sec:additional_pop_trans_ex}

\setcounter{equation}{0}
\setcounter{figure}{0}
\setcounter{table}{0}
\renewcommand{\theequation}{B\arabic{equation}}
\renewcommand{\thefigure}{B\arabic{figure}}
\renewcommand{\thetable}{B\arabic{table}}

\begin{figure}[h]
    \centering
    \subfloat[][Histograms of immune status]{\includegraphics[width=0.9\textwidth]{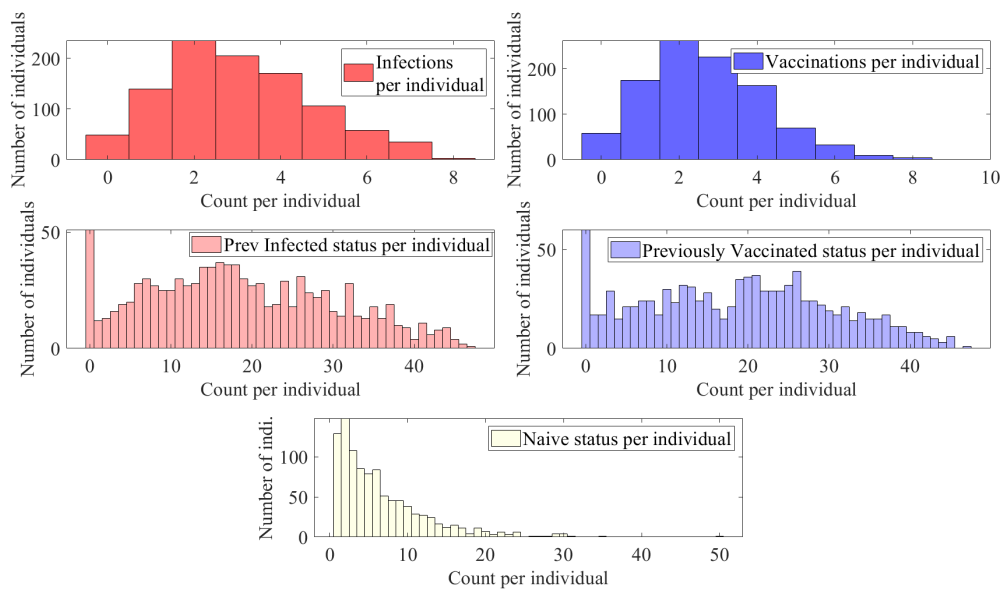} \label{fig:S3_counts}} \\
    \subfloat[][Five representative individual sequences]{\includegraphics[width=0.9\textwidth]{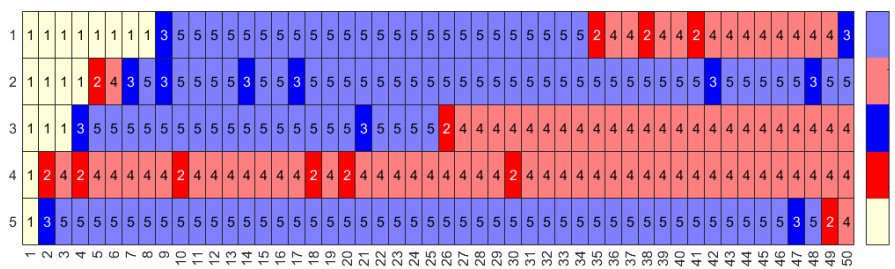} \label{fig:S3_sequences}}
    \caption{Simulations for a population with $1000$ individuals with a time-homogeneous transition matrix $S_3$, per $50$ time steps, i.e., approximately two year period. (a) Histograms of counts of total infections, vaccinations, previously infected status, previously vaccinated status, and na{\"i}ve status.
 Five representative sequences 
 Light yellow denotes the na{\"i}ve state (1, $N$), red and pink  the newly and previously infected states (2, $\pos$ and 4, $\posp$), and blue and light blue the newly and previously vaccinated states (3, $\vac$ and 5, $\vacp$).}
\end{figure}

Let us now consider $S_3$, which is obtained from $S_1$ by considering a population with comparably high rates of infection and vaccination.  

\begin{equation}
    S_3 = \TM{0.1}{0.1}{0.08}{0.05}{0.05}{0.06}{0.01}{0.01}
\end{equation}
 As expected, the number of vaccinations per individual and stay in previously vaccinated states rises for $S_3$ as compared to $S_2$; see Figure \ref{fig:S3_counts}. Moreover, the time spent in na{\"i}ve or na{\"i}ve-like states decreases. We see that individuals frequently move from a previously infected/vaccinated state to another immune event; see Figure \ref{fig:S3_sequences}.

\section{Additional personal trajectories of interest}
\label{sec:add_pers_traj_ex}

\setcounter{equation}{0}
\setcounter{figure}{0}
\setcounter{table}{0}
\renewcommand{\theequation}{C\arabic{equation}}
\renewcommand{\thefigure}{C\arabic{figure}}
\renewcommand{\thetable}{C\arabic{table}}

\begin{figure}[h]
\centering
\includegraphics[width=1.0\linewidth]{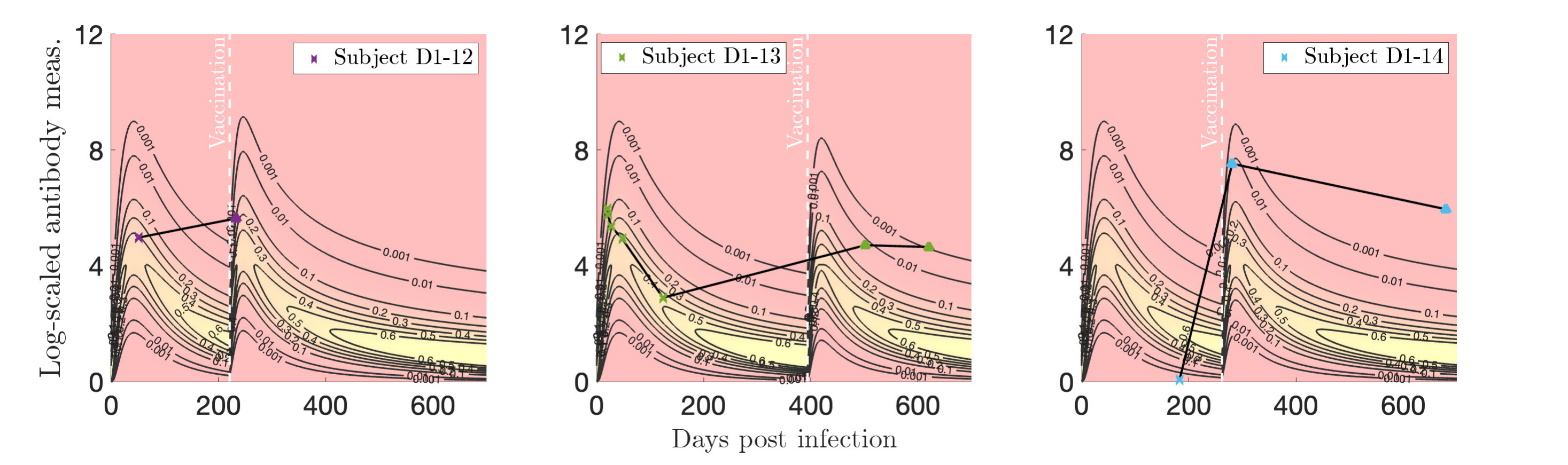}
\caption{Log-transformed antibody measurements from the infected then vaccinated (VI) population from Dataset 1 with personal trajectories of interest with corresponding probability models. The relative date of vaccination is marked with a vertical, white dashed line.}
\label{fig:UVA_traj_interest_inf_vax}
\end{figure}

 In Figure  \ref{fig:UVA_traj_interest_inf_vax} we plot individualized two-event models for vaccination after infection for three subjects of interest. Although the models shown in  Section  \ref{sec:results_data} take into account the time between an event and the day of measurement, it would be difficult to visualize this time span in the two-event figures, as there is often no associated measurement value for the day of the event and the extra information may crowd already densely packed data points.  When studying personal trajectories, one option is to mark the line connecting available measurements with an additional marker symbolizing the day of the event. Here, while studying personal trajectories one by one, we mark the date of vaccination (after infection or as a booster dose) as a vertical, dashed line. Subject D1-12 has a representative relative time between infection and vaccination (182 days), and exhibits a standard increase in antibody response after vaccination. Both measurement values fall on the higher side of observed values. Subject D1-13 exhibits a strong initial response to infection and a typical decay. Nearly 400 days elapse between their infection and vaccination; likely as a result, they show a strong and sustained response to vaccination. Subject D1-14 is an outlier for their infection response, displaying a very low antibody level almost 200 days after infection, but perhaps in response, they exhibit a very strong increase in antibody level after vaccination.

 \section{When to measure the event of  vaccination?}
 \label{sec:dose1_or_dose2}
 
 \setcounter{equation}{0}
\setcounter{figure}{0}
\setcounter{table}{0}
\renewcommand{\theequation}{D\arabic{equation}}
\renewcommand{\thefigure}{D\arabic{figure}}
\renewcommand{\thetable}{D\arabic{table}}

\begin{figure}[h]
    \centering
\includegraphics[width=0.8\linewidth]{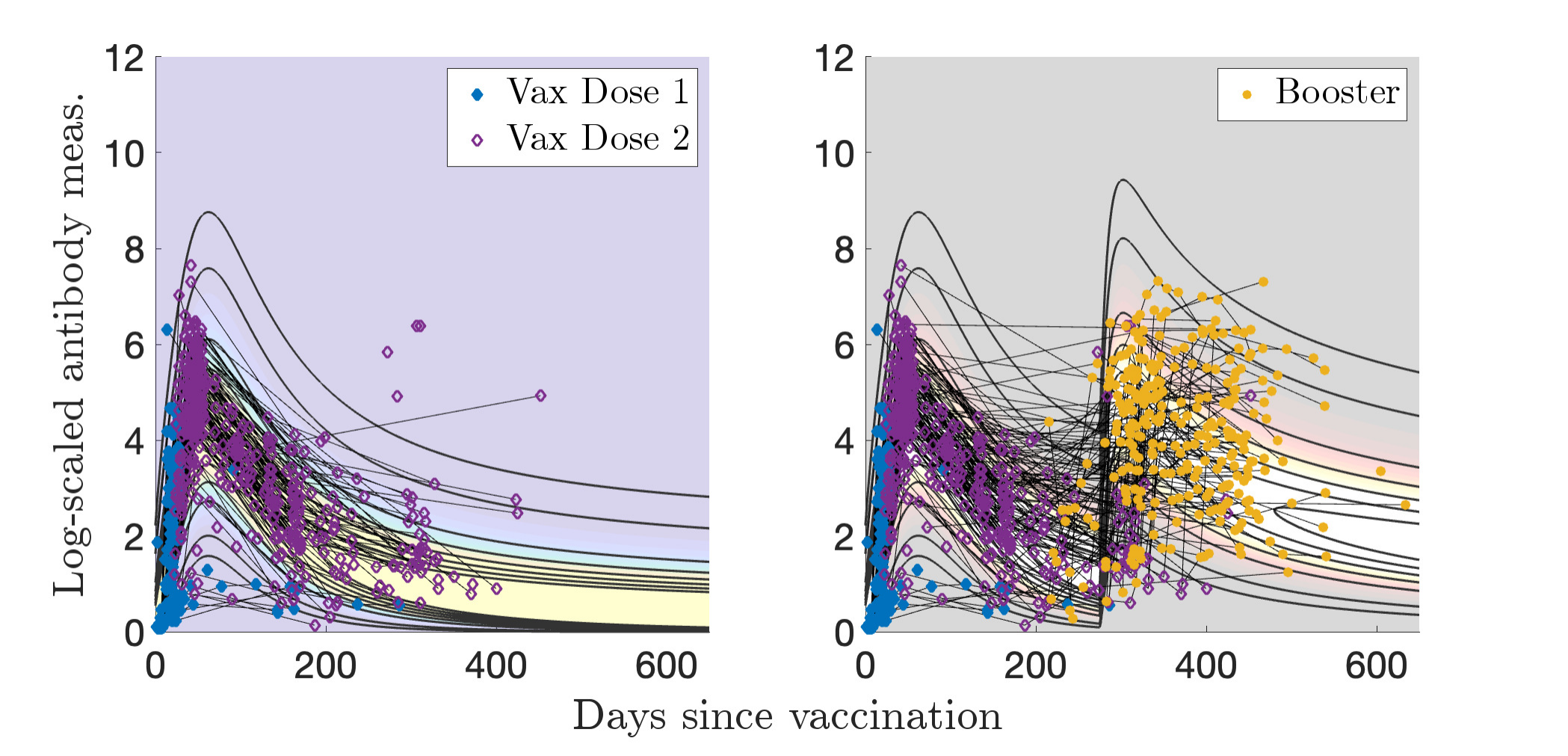}
    \caption{Log-transformed antibody measurements from the vaccinated (V) and boosted (VV) populations from Dataset 1 with corresponding probability models. }
    \label{fig:UVA_models_k_vax_dose1}
\end{figure}

Because some vaccines are given as a multi-dose sequence, and an individual is not considered fully vaccinated until the sequence is finished, it is interesting to study the effect on our models of setting $t = 0$ to represent the day of the first dose of vaccination, rather than the second, as we have done in  Section  \ref{sec:results_data}.
 The relative days between the first and second dose of vaccination were available for Dataset 1. We redid our models using dose 1 as $t = 0$ (call this Vax Model Dose 1); the results are shown in Table \ref{table:MLE_params_UVA_dose1} and Figure \ref{fig:UVA_models_k_vax_dose1}. We note that this allows for more data to be used; in   Section \ref{sec:results_data}, only measurements post dose 2 are relevant for the model (call this Vax Model Dose 2; see  Figure \ref{fig:UVA_models_k_traj}).

 Compared to Vax Model Dose 2, $\theta_{\vac_1}$ and $\phi_{\vac_1}$ are an order of magnitude smaller for Vax Model Dose 1; as a result, the peak and decay of Vax Model Dose 1 are smaller and start later, around 60 days as opposed to 10. Ultimately, we are observing the same peak in the two models, which is the response to the second immunization.  More of the increase in antibody response is visible when including data after only the first dose in the model; this is paralleled by the infection model (see  Figure \ref{fig:UVA_models_k_traj}). Perhaps reassuringly, this modeling choice for the first vaccination event has minimal effect on the model for the booster dose; the parameter values are relatively similar for Vax Model Dose 1 and Vax Model Dose 2.

\begin{table}[h]
 \centering
\begin{tabular}{|l|r|r|r|r|r|r|r|r|r|r|}
\hline
\multicolumn{1}{|c|}{\textbf{Data}} & \multicolumn{1}{c|}{$\bm{\theta_{\vac_1}}$} & \multicolumn{1}{c|}{$\bm{\phi_{\vac_1}}$} & \multicolumn{1}{c|}{$\bm{k_{\vac_1}}$} & \multicolumn{1}{c|}{$\bm{\theta_{\vac_2}}$} & \multicolumn{1}{c|}{$\bm{\phi_{\vac_2}}$} & \multicolumn{1}{c|}{$\bm{k_{\vac_2}}$} \\ \hline
Vaccination & 44.1 & 2.17 & 2.56 &  &  &  \\ \hline
\begin{tabular}[c]{@{}l@{}}Vaccination, \\ booster \end{tabular} & {\color[HTML]{9B9B9B} 44.1} & {\color[HTML]{9B9B9B} 2.17} & {\color[HTML]{9B9B9B} 2.56} & 155 & 12.6 & 1.51 \\ \hline
\end{tabular}
\caption{Optimal parameters from maximum likelihood estimation for Dataset 1 for measurements corresponding to vaccination as the first event. Grayed out numbers indicate variables that are part of the model but are not part of the optimization for that data. Time is measured since first vaccination dose.}
\label{table:MLE_params_UVA_dose1}
\end{table}

\FloatBarrier

\section{Supplementary figure captions}
\label{sec:supp_video_captions}

\paragraph{Supplementary Figure 1 (GIF).} Log-transformed antibody measurements from the vaccinated then boosted (VV) population from Dataset 1 with personal trajectories of interest with corresponding probability models. The relative date of the booster dose is marked with a vertical, white dashed line. See Figure \ref{fig:UVA_traj_interest_vax_vax} for accompanying static example images.

\paragraph{Supplementary Figure 2 (GIF).} Log-transformed antibody measurements from the infected, then vaccinated (IV) population from Dataset 2 with personal trajectories of interest with corresponding probability models. The relative date of the vaccination dose is marked with a vertical, white dashed line. Measurements with recorded monotonically decreasing OD values have been AUC-extrapolated using the next titration level. See Figure \ref{fig:UCSD_inf_vax_traj_interest} for accompanying static example images.

\end{document}